\documentclass[%
apsrev,
prb,
preprint,showkeys%
]{revtex4-1}
\usepackage{graphicx}
\usepackage{dcolumn}
\usepackage{bm}
\usepackage{physics} 
\usepackage[utf8]{inputenc}
\usepackage[T1]{fontenc}
\usepackage{mathptmx}
\usepackage{etoolbox}
\usepackage{xcolor}
\usepackage{amsmath}
\usepackage{amssymb}
\usepackage{caption}
\usepackage{subcaption}
\usepackage{float}
\usepackage{tikz}
\usepackage[toc,page]{appendix}
\usepackage{hyperref}
\usepackage{graphicx,booktabs,array}
\usepackage{stackengine}

\newcommand{\dbr}{\text{d}\mathbf{r}}

\newcommand{\br}{\mathbf{r}}
\newcommand{\dm}{\rho_{1}}
\newcommand{\dmrrp}{\dm(\br, \br')}
\newcommand{\dmxxp}{\dm(x, x')}

\newcommand{\gfrrp}{G(\mathbf{r}, \mathbf{r}'; \beta)}

\newcommand{\gfrpr}{G(\mathbf{r}', \mathbf{r}; \beta)}

\newcommand{\gfxxp}{G(x,x'; \beta)}

\newcommand{\ham}{\hat{\mathcal{H}}}

\newcommand{\ef}{\epsilon_F}

\begin{document}
\title{ A new Green's function formalism for kinetic energy density functional for atomic and molecular system: Emergence of $N-$dependence using model potentials}
\author{Priya}
\affiliation{Department of Chemistry, Indian Institute of Technology Kanpur, Uttar Pradesh, India}
\author{Mainak Sadhukhan}
\email{mainaks@iitk.ac.in}
\affiliation{Department of Chemistry, Indian Institute of Technology Kanpur, Uttar Pradesh, India}

\begin{abstract}
    An accurate expression of the kinetic energy density of an electronic distribution in terms of the single particle reduced density matrix for atomic and molecular systems is a long-standing problem in electron structure theory. Existing kinetic energy density functionals are generally expressed as modifications over kinetic energy of homogeneous electron gas and/or von Weizs\"acker  kinetic energy. A large class of these functionals also require empirical parametrizations to make accurate predictions of the kinetic energy for atomic and molecular systems restricting their transferability. Moreover, the correct kinetic energy density which produces accurate local properties such as atomic shell structure is still an unsolved problem. In this work, we have developed an exact methodology that can be used to derive the kinetic energy of an electronic system of arbitrary spin multiplicity. One of the attractive features of this present analytical formalism is the possibility of systematic improvement of the kinetic energy by virtue of a novel perturbation series. Applying this methodology to simple model systems such as one-dimensional quantum harmonic oscillator and homogeneous electron gas produces a qualitatively correct $N$-dependence of kinetic energy as a result. A one-to-one correspondence between our formalism to the traditional Green's function formalism is also demonstrated.
   
\end{abstract}
\keywords{Orbital-free density functional theory; Kinetic energy density; Green's function ; Weak interaction potential }

\maketitle

\section{Introduction}\label{intro}
An accurate approximation of kinetic energy (KE) functional with correct local behavior is the main bottleneck in the development of orbital-free density functional theory (OF-DFT)\cite{Ludena2002,ofdft_2013wang,David_PhysRevA,Perdew_PhysRevB,GarciaCervera2008,LucaPhysRevB,Eek2006266,Karasiev2006111,karasiev2009PhysRevB,Trickey20092943,carter2010PhysRevB,KARASIEV20122519}. In this work, we have developed a general, spin-multiplicity-independent perturbative formalism to systematically calculate the KE density (KED) for atomic and molecular electron densities in their ground electronic state. Our method employs a novel series-expansion technique of Green's function, which asymptotically produces Thomas-Fermi and von Weizs\"acker KE ($T_{vW}[\rho]$) as limits. In this work, we have also shown the $N-$dependence of kinetic energy for one-dimensional model potentials as proof-of-concepts. However, our formalism is not limited to these flagship model systems and can be readily extended to real systems. \\

 The challenge of finding an accurate approximation of KED is related to its definition in terms of the one-particle density matrix (ODM) $\dmrrp$. The kinetic energy and the KED $t(\br, \rho(\br))$ of an electronic system are defined  by (atomic units have been used throughout the article unless mentioned explicitly)
 \begin{equation}\label{tsdef}
     T[\rho] = \int t(\br, \rho(\br)) \dbr
 \end{equation}
 where
\begin{equation}
    t(\br, \rho(\br)) =-\frac{1}{2}\nabla^2 \dmrrp\vert_{\br=\br'}.
\end{equation}
Existing methods to compute the KED start with approximations of $\dmrrp$\cite{ParrYang1994}.
 To understand the importance of finding a proper KED lies in the origin of density functional theory (DFT) itself. One particle reduced density matrix $\rho(\br)$ is the fundamental variable in DFT.  $\rho(\br)$ is a three-dimensional quantity irrespective of system size that can be obtained from a single Euler-Lagrange equation\cite{HohenKohn1964}
 {\begin{equation}\label{euler}
     \frac{\delta E[\rho]}{\delta \rho} = \mu
  \end{equation}}
where $E[\rho]$ and $\mu$ are total electronic energy as a functional of $\rho(\br)$ and chemical potential respectively. The functional form of total energy $E[\rho]$ is unknown making the exact solution of Eq.\eqref{euler} an intractable problem. The $E[\rho]$ is the sum of potential energy $V[\rho]$ and kinetic energy $T[\rho]$. In contrast to $V[\rho]$, for which adequately accurate approximations exist for atomic and molecular systems, similar approximations are yet to be obtained for $T[\rho]$. Apart from an accurate kinetic energy value, $\frac{\delta T[\rho]}{\delta \rho}$ should also produce local characteristic features of the electronic system such as shell structure for atoms and Friedel oscillations for bulk solids.  Kohn-Sham DFT (KS-DFT) \cite{kohnsham1965} bypasses this challenge by mapping the interacting system to a system of $N_{ni}$ non-interacting fictitious particles. The shell structure arises from the Pauli exclusion principle and is already implicit in orbital-based approaches like KSDFT. The KS-KE $(T_{s}[\rho])$ is defined as $T_{s} [\rho]= \sum_{i}^{N_{ni}} T_{i}$, where $T_{i}$ is the KE of the $i^{th}$  non-interacting particle. The Pauli potential $v_P(\rho(\br))= \frac{\delta T_{P}[\rho]}{\delta \rho} = \frac{\delta(T_{s}[\rho] - T_{vW}[\rho])}{\delta \rho}$ has been shown to account for the shell structure \cite{Tal1978153,Bartolotti19824576,Levy1988625}. One of the leading approximations of $v_P$ employs bifunctional formalism and is used in computing molecular Pauli potentials \cite{finzel2021molecule} as well as in quantum crystallography \cite{finzel2021detailed}. Moreover, the dependence of KE  on the number of electrons $N$ and atomic charge $Z$ is also an open-ended problem. Furthermore, the computational cost of KS-DFT increases significantly with $N_{ni}$ due to the need to solve $N_{ni}$  coupled differential equations. In that context, OF-DFT provides a lucrative alternative.  \\
The earliest approximation of KE was modeled after homogeneous electron gas (HEG) \cite{thomas1927calculation, fermi1927statistical} 
\begin{equation}
    T_{TF}[\rho] = C_{TF}\int  \rho^{5/3}(\br)d\br; \, C_{TF} = \left(\frac{3}{10}\right)\left(3 \pi^2\right)^{2/3}
\end{equation}
followed by von Weizs\"acker's correction for slightly inhomogeneous electron density \cite{weizsacker1935theory}
\begin{equation}
    T_{vW} [\rho] = \frac{1}{8}\int  \frac{\abs{\nabla \rho(\br)}^2}{\rho(\br)} d\br.
\end{equation}
 Note that the $T_{vW}$ is the exact kinetic energy for electrons in $1s$ orbital of a hydrogen atom. These functionals are inadequate to describe molecular binding \cite{Teller_RevModPhys.34.627}. Generally, $T_{TF}$ augmented by a scaled von Weis\"acker term in the form 
\begin{equation}
    T[\rho] = T_{TF}[\rho] + \lambda T_{vW}[\rho]
\end{equation}
has been considered as generalised Thomas-Fermi-von Weizs\"acker (TF$\lambda$W) kinetic energy. $\lambda=1$ overestimates the atomic KE.  It has been found empirically that $\lambda = 1/5$ is the best fit for Hartree-Fock electron density. Analytical derivation of $\lambda$ using generalized gradient approximations (GGA), on the other hand, predicts $\lambda=1/9$.
Additionally, no shell structure is obtained in electron density using TF$\lambda$W theory. As a consequence, standalone quantum chemical techniques could not be developed based on this approach \cite{Wang_PhysRevB.45.13196}. Use of the $T_{TF}$ as the zero$^{th}$ order approximation is the suspect here. The density profile of atoms are more closely represented by a hydrogen atom than HEG. As a result, the exact KE is qualitatively closer to $T_{vW}$ than $T_{TF}$. 
In a series of studies for closed-shell systems\cite{Alonso1978,Alonso1981,debghosh1983},  the correlational function $C(\br,\br') = \frac{\dmrrp}{\rho(\br) \rho(\br')}-1 $ has also been used as the starting point of further approximations. Such consideration leads a KE 
\begin{equation}\label{t_debghosh}
    T[\rho] = T_{vW}[\rho] + C_{TF}\int f(\br) \rho^{5/3}(\br) d\br.
\end{equation}
The enhancement factor $f(\br)$ is an unknown function of $\rho(\br)$. It also has been argued that $f(\br)$ contains the information about atomic shell structures\cite{debghosh1983}. This form of KE has been successfully employed in developing an amalgamation of quantum fluid dynamics and DFT\cite{debghosh1982_jcp,deb1989, AKRoy1999}. The resultant Deb-Chattaraj equation has been used to compute the response of electron densities of atoms and molecules in the presence and absence of external fields \cite{akroyrsinghdeb_1997, singh_deb_96,wadehra_deb_2006, sadhukhandeb_2014}.  \\
In the condensed matter physics and materials science community, the kinetic energy density functional (KEDF) is enhanced using non-local functionals to improve upon the Thomas-Fermi (TF) and von Weizsäcker (vW) terms. The Wang-Govind-Carter (WGC) functional is a non-local KEDF that incorporates local density-dependent terms and gradient correction. It is defined as $T_{WGC}[\rho] = T_{TF}[\rho]+ T_{vW}[\rho]+T_{NL}[\rho]$ where $T_{NL}$ is called the non-local KE
 \cite{carter_2018, carter_wang1999orbital, Carter_Ligneres2005, Carter_HUNG2009163}. Non-local functionals, such as generalized gradient approximation (GGA)-type functionals and density-dependent kernels, have been considered \cite{Wang_teter_PhysRevB.45.13196, carter_PRB2010, Pavanello_JCP_10.1063/1.5023926, Pavanello_PRB2021, Carter_Ligneres2005, Carter_2018_OFDFTmaterialresearch}. To improve the accuracy of existing functionals, methods using line integrals \cite{XWM_PhysRevB.100.205132} and/or the Perdew-Burke-Ernzerhof (PBE) exchange like enhancement factor  \cite{revHC_PhysRevB.104.045118} have been introduced.  
A number of computational packages \cite{carter_profess,gpaw_ofdft_14} are also available. Recently, significant work has been done using machine learning to find KEDFs for solids, atomic, and molecular systems \cite{ML_Snyder_PRL2012,mayeretal2020,sun_MLPhysRevB.109.115135,ML_JCP_2018,ML_JCP_Jacob2018}. In contrast to the bulk solid systems, the electron density is more inhomogeneous for atoms and molecules. As a result, the enhancement factors applicable for solids may not be useful for atoms and molecules.  In our recent work \cite{Priya_doi:10.1080/00268976.2022.2136114}, we tested this hypothesis \cite{Constantin2019} and confirmed this conclusion. \\
Apart from being specific model-dependent these previous works also fail to explain how $T[\rho]$ and $t(\br, \rho(\br))$ are dependent on $N$, $Z$, and the spin multiplicity of the system. Also, the accuracy of the approximations involved were bench-marked against the global (integrated) value of $T[\rho]$, which fudges local details of the $t(\br, \rho(\br))$. Note that these local features are quintessential to produce the correct shell structure. Therefore, a systematic, specific model-independent methodology is required which should be able to explain $N$ and $Z$ dependency and can be employed for arbitrary spin-multiplicity.\\
In our recent work \cite{Priya_doi:10.1080/00268976.2022.2136114}, we developed a systematic methodology to find kinetic energy density, using Green's function (GF) which satisfies all aforementioned qualities. We will call it Green's function formalism (GFF). The GF employed in this present work is defined as the Laplace transform of the ODM
\begin{equation}\label{gfdef}
    \gfrrp = \int_{0}^{\infty} \text{d}\ef e^{-\beta \ef} \dmrrp \equiv \bra{\br}e^{-\beta \ham }\ket{\br'}.
 \end{equation}
Here $\ef$ and $\ham$ are Fermi energy and Hamiltonian of the system respectively. The auxiliary variable $\beta$ is mathematically similar to the inverse temperature used in Statistical mechanics. However, in this formalism, the similarity is only notational and does not signify any physical thermodynamic temperature. The present formalism is a $0$ K theory. The ODM can be retrieved from the GF via inverse Laplace transform \cite{ParrYang1994}. 
\begin{equation}\label{bromwich}
  \dm(\br,\br') = \lim_{\Gamma \to \infty}\frac{1}{2\pi i} \int_{\gamma-i\Gamma}^{\gamma+i\Gamma}\frac{\text{d}\beta}{\beta}e^{\beta \ef} \gfrrp
 \end{equation}
 where $\gamma \in \mathbb{R}_{>} $.
In this work, we have developed a systematic expansion technique for the GF applicable for atomic and molecular electron densities akin to perturbation theory. \\
After a brief introduction of existing approaches in Section \ref{intro}, we formulate our methodology in Section \ref{gff}. This exposition is followed by a discussion regarding the connection between traditional Green’s functions\cite{economou2006green} (GF) and our Green’s function formalism (GFF) in Section \ref{tradgf}. To demonstrate the efficacy of the GFF methodology, we applied it to simple model systems, such as a one-dimensional harmonic oscillator in the weak interaction limit in Section \ref{gfweakfield} followed by a comparison with the results from traditional GF in Section \ref{tradgf_1dsho}. After analyzing the spatial nature of the GF in Section \ref{gfnature} we analyze the dependence of kinetic energy on the number of electrons ($N$) and the Fermi energy ($\epsilon_{F}$) in Section \ref{nzdep}. After testing the methodology on noble gas atoms (Section \ref{noblegas}) we show how a limit of the weak interaction potential ($W_{0}$) naturally arises from our formalism in Section \ref{weakfieldlim}. The paper concludes with Section \ref{conclude}.

\section{Analytical formalism} \label{Theory}
\subsection{Green's function formalism}\label{gff}
 To begin our Green's function formalism (GFF), we define the atomic Hamiltonian 
 \begin{equation}\label{ham_partition}
     \ham = \underbrace{\ham_0+\ham_{Z}}_{\ham_{H}} + \hat W,
 \end{equation}
where $\ham_0$, $\ham_{Z}$, and $\hat W$ are free-particle Hamiltonian, electron-nucleus attraction potential and inter-electronic repulsion potentials, respectively. The Hydrogenic Hamiltonian with atomic charge $Z$ is defined as $\ham_{H} = \ham_0+\ham_{Z}$. The corresponding GFs are   
\begin{equation}\label{ghdef}
 G^{H}(\br, \br'; \beta) = \bra{\br}e^{-\beta \ham_{H}}\ket{\br'}
\end{equation}
and
\begin{equation}
 G^{0}(\br, \br'; \beta) = \bra{\br}e^{-\beta \ham_{0}}\ket{\br'},
\end{equation}
 respectively. For atomic systems, the contribution of the inter-electronic repulsion $\hat{W}$ is smaller compared to $\ham_{H}$ (\emph{vide} Eq.\eqref{ham_partition})  \footnote{Consider Helium and Lithium atoms with exact non-relativistic energies as $\approx -2.904$ \textit{a.u.} and $\approx -14.864$ \textit{a.u.}, respectively. To them the hydrogenic parts of the Hamiltonian attribute $\approx -4.0$ \textit{a.u.} and $\approx -20.25$ \textit{a.u.} The contributions of the inter-electronic repulsion are evidently smaller compared to the hydrogenic part.}. As a result, $G^{H}(\br, \br'; \beta)$ is considered as the natural leading order term of $G(\br, \br'; \beta)$. This is in contrast with previous attempts where the leading order term was chosen from an arbitrary model system \cite{yang1986gradient,Brack1976}. HEG Hamiltonian was almost exclusively used as this leading order Hamiltonian despite the fact that the atomic and molecular systems are qualitatively very different from HEG. In Section \ref{intro}, we have presented the true kinetic energy of an atom as the admixture of two extreme limits viz. HEG ($T_{TF}$) and von Weizs\"acker ($T_{vW}$) kinetic energies. The $T_{vW}$ represents a hypothetical Bosonic hydrogenic system where all particles occupy the ground state of the atom \emph{viz} $1s$ orbital. Note that the hydrogenic Green's function $(G^{H})$ employed in our approach (Eq.\eqref{ghdef}) is of Fermionic nature and therefore yields kinetic energy (say $T_{H}$) which is different from the $T_{vW}$. We can relate $T_{H}$ and $T_{vW}$ in the following manner. Consider a Bosonic system where all the population is on the ground state. On the other hand, the Fermions follow the Pauli exclusion principle leading to exchange-hole. The exchange-hole is responsible for pushing the electrons to higher energy states\cite{dyson_lenard_67,lioeb_thirring_75}. If we switch off the exchange-hole, all electrons occupy the ground state $1s$. As a result, $T_{H}$ become $T_{vW}$. Since the kinetic energy of the excited states is higher than that of the ground state, the latter represents a part of the former \emph{ie} $T_{H}> T_{vW}$. 
 
 On the same note, we can interpret the interaction Green's function $(G^{int})$ defined as the difference between total Green's function $(G)$ and the hydrogenic Green's function $(G^{H})$ 
 \begin{equation}\label{g_rrp_rep}
   G^{int}(\br, \br'; \beta) \equiv \bra{\br}\hat O\ket{\br'} = G(\br, \br'; \beta)- G^{H}(\br, \br'; \beta).
 \end{equation}
 Here 
 \begin{equation}
     \hat O = e^{-\beta \ham} - e^{-\beta \ham_{H}}.
 \end{equation}
 Since $G^{int}$ contains inter-electronic repulsion, this term represents the central challenge for our formalism.

  In order to get the systematic expansion of $G^{int}$ we expand $\hat{O}$ using Zassenhaus formula \cite{CasasNadinic2012, magnus1954} as
 \begin{equation}\label{hato_zassenhaus}
     \hat O = - e^{-\beta \ham_{H}}\left( 1-e^{-\beta \hat W} e^{-\frac{\beta^2}{2} \comm{\ham_H}{\hat W}} \ldots \right).
 \end{equation}
We find
\begin{equation}\label{comm_H}
    \comm{\ham_H}{\hat W}=\comm{\ham_0}{\hat W} 
\end{equation}
since $\comm{\ham_Z}{\hat W}$ commutes. However, such replacements cannot be done for terms containing triple commutators onwards, leading to our first approximation
\begin{widetext}
   \begin{equation}\label{hatohydr_heg}
     \hat O \approx e^{-\beta \ham_{H}} e^{\beta \ham_{0}}\left(e^{-\beta (\ham_0+\hat W)} - e^{-\beta \ham_{0}}\right) .
\end{equation} 
\end{widetext}
Note that despite this approximation in the third-order commutator onwards, the expansion of the second-order commutator represents a partial sum up to the infinite order interaction. Therefore it is of the similar spirit of random phase approximation \cite{gellmann_bruckner_57}. Additionally, this approximation is qualitatively acceptable where $\ham_0 \approx \ham_H$,  away from the nucleus where $\ham_Z \to 0$. Note that we have not completely replaced $\ham_{Z}$ by $\ham_{0}$ since the leading term of the expansion is still $G^{H}(\br, \br', \beta)$. Also, this approximation is consistent with the fact that $f(\br) \to 1$ away from the nucleus. We found, however, that this approximation leads to an unphysical asymmetry in the resulting GF with respect to $\br$ and $\br'$. 
To minimize the effect of asymmetry we decided to use a symmetrized GF
\begin{equation}\label{symgf}
    {G}(\br, \br';\beta) \equiv \frac{1}{2}\left[\gfrrp+\gfrpr\right]
\end{equation}
as our fundamental GF where $\gfrrp$ in the RHS of the above equation is defined as 
\begin{widetext}
 \begin{equation}\label{resoid}
    \gfrrp = {G^{H}}(\br, \br'; \beta)+ \int \dbr'' \bra{\br}e^{-\beta \ham_{H}}\ket{\br''}\bra{\br''}\sum_{n=0}^{\infty} \frac{\beta^n}{n!}\ham_{0}^{n}\left(e^{-\beta (\ham_0+\hat W)} - e^{-\beta \ham_{0}}\right) \ket{\br'}.
 \end{equation}   
\end{widetext}
Eqs.\eqref{symgf} and \eqref{resoid}, taken together, are the governing equations to systematically calculate the KED.  The infinite sum over $n$ in Eq.\eqref{resoid} represents a convergent series of terms thereby facilitating a systematic scheme of approximations. This feature of our formalism allows for a rigorous development scheme without re-coursing to any parameter tuning. Each successive approximation will provide greater accuracy in the similar spirit of ``Jacob's ladder'' in XC functionals. In our scheme, however, no empirical parameter would be required.  

\subsection{Connection to traditional Green's function}\label{tradgf}

In  GFF we have used Eq.\eqref{gfdef} as the basis of our methodology. This form of Green's function can be shown to be related to more traditional version of Green's function\cite{economou2006green} defined as 
\begin{equation}\label{deftradgf}
    \tilde{G}(\br, \br'; E) = \bra{\br}\frac{1}{\ham- E}\ket{\br'}.
\end{equation}
 Numerous methods have been developed to compute $\tilde{G}$ in the problems of quantum field theory, statistical mechanics, and related fields. As a result a mathematical connection between $G$ and $\tilde{G}$ will allow for direct comparison as well as the prospect of employing hitherto existing methods in our formalism. For example, $\tilde{G}$ can be expanded in terms of the Dyson equation with the well-understood definition of self-energy $\Sigma$. Highly effective techniques based on Feynman diagrams have been developed to compute $\Sigma$. The connection can be achieved by employing an auxiliary variable $\mu$  \emph{via} 
\begin{equation}
    \gfrrp = \bra{\br}e^{-\beta \ham}\ket{\br'} = \lim_{\mu \to 1^+} \bra{\br}e^{-\mu\beta \ham}\theta(\mu)\ket{\br'}
\end{equation}
where $\theta(\mu)$ is Heaviside theta function.
A subsequent introduction of Fourier transform $\hat{\mathcal{F}}_{\mu}^{\omega}$ over $\mu$ produces (see Appendix \ref{gtogtilde} for details) the desired relation:
\begin{equation}
    \gfrrp = \lim_{\mu \to 1^+} \frac{1}{\beta} \hat{\mathcal{F}}_{\mu}^{\omega}( \tilde{G}(\br, \br'; \tau)).
\end{equation}    
The Fourier operator $\hat{\mathcal{F}}_{\mu}^{\omega}$ converts $\mu$ space functions to $\omega$ space functions and ${ \tau = -i \frac{\omega}{\beta}}$. Since the $\lim_{\mu \to 1^+}{} $ and Fourier transform are both linear operators once can expand $\tilde{G}(\br, \br'; \tau)$ perturbatively before employing $\lim_{\mu\to1^+}$ and $\hat{\mathcal{F}}_{\mu}^{\omega}$, keeping their relative ordering intact. The operator $\hat{\tilde{G}}$ can be expanded through Dyson equation
\begin{equation}\label{dyson}
   \hat{ \tilde{G}} = \hat{\mathcal{G}}_{0} +\hat{\mathcal{G}}_{0} \hat{W} \hat{\mathcal{G}}_{0} +
   \hat{\mathcal{G}}_{0} \hat{W}\hat{\mathcal{G}}_{0} \hat{W}\hat{\mathcal{G}}_{0}  + \ldots .
\end{equation}
Here $\hat{\mathcal{G}}_{0}$ is the GF operator for the unperturbed Hamiltonian which yields $\mathcal{G}_{0}(\br, \br'; \tau) $ in  position representation and $\hat{W}$ is the mean field interaction potential. The unperturbed Hamiltonian is $\hat{H}_{H}$ in our case prompting  $ \mathcal{G}_{0}(\br, \br'; \tau) \equiv \mathcal{G}^{H}(\br, \br'; \tau)$. The integral form of the Dyson equation can be written as 
\begin{equation}\label{dysongf}
  \begin{split}  
    \tilde{G}(\br,\br';\tau) =\mathcal{G}^{H}(\br,\br';\tau) \\+ \int\int d\br'' d\br''' \mathcal{G}^{H}(\br,\br'';\tau)W(\br'', \br''')\tilde{G}(\br''',\br',\tau).
  \end{split}    
\end{equation}
In this work, $\hat{W}$ is considered as a mean field interaction potential prompting  $W(\br'', \br''') \equiv \delta(\br''-\br''') W(\br'')$. As a result, Eq.\eqref{dysongf} becomes
\begin{equation}\label{traditional_gf}
  \begin{split}  
    \tilde{G}(\br,\br';\tau) = \mathcal{G}^{H}(\br,\br';\tau) \\ + \int d\br'' \mathcal{G}^{H}(\br,\br'';\tau)W(\br'')\tilde{G}(\br'',\br',\tau).
  \end{split}    
\end{equation}
The expansion for $\tilde{G}$ also has $\mathcal{G}^{H}$ as the leading order term similar to our GFF while the second term represents a parallel to $G^{int}$. 

\section{Results and Discussion} \label{results}
\subsection{Weak interaction limit for one-dimensional simple harmonic oscillator}\label{gfweakfield}
 To exemplify the efficacy of our GFF we apply it to a one-dimensional simple harmonic oscillator. Also, we have refrained from using the atomic units in the following sections to clarify the relative magnitudes of different terms. The ground state of this system with natural frequency $\omega$ and mass $m$ is described by the probability density 
 \begin{equation}\label{rhosho}
     \rho^{SHO}(x) = \left(\frac{m\omega}{\pi\hslash}\right)^{1/2}e^{-\frac{m\omega x^2}{\hslash}}
 \end{equation}
 where $\omega$ represents the inverse width of the probability distribution. As a result, larger $\omega$ leads to a faster-decaying probability density from the center \emph{i.e.} $x=0$.  
The corresponding GF is
\begin{widetext}
    \begin{equation} \label{GFmain}
   \gfxxp = {G}^{SHO}(x, x'; \beta)+ \int dx'' {G}^{SHO}(x, x''; \beta) \bra{x''}\sum_{n=0}^{\infty} \frac{\beta^n}{n!}\ham_{0}^{n}\left(e^{-\beta (\ham_0+\hat W)} - e^{-\beta \ham_{0}}\right) \ket{x'}.  
\end{equation}
\end{widetext}
In this equation, $\hat{W}$ is the mean-field interaction potential. 
 Note that the nuclear charge ($Z$) in an atom signifies the extent of force that pulls electrons towards the nucleus in the center. For a harmonic oscillator, a similar role is played by the force constant $k = m\omega^2$. Therefore Eq.\eqref{resoid} and Eq.\eqref{gf_shofree} are qualitatively equivalent except that for the latter only bound states are available.
Furthermore, we have considered the weak-field limit since it resembles the low-density limit of the Gell-Mann-Bruckner model\cite{gellmann_bruckner_57} where particles move under a constant positive background leading to constant charge density. As a result, the mean-field interaction potential also varies negligibly over space. To accomplish this, we have expanded $\hat{W}$ in power series 
\begin{equation}\label{weakfield}
     W(x) = W_{0} + x \partial_{x} W(x)|_{0}+ \hdots 
 \end{equation}
and considered only the first constant term $W_{0}$.  This approximation is equivalent to $\ham_0 \approx \ham_H$. 
By virtue of this approximation, Eq.\eqref{GFmain} becomes
\begin{widetext}
\begin{equation}\label{gf_shofree}
 \gfxxp = {G}^{SHO} (x, x' ; \beta) + \int dx'' (e^{-\beta  W_{0}}-1 ) \left(    {G}^{SHO}(x, x'' ; \beta) e^{-\beta \frac{\partial}{\partial\beta}} {G}^{0}(x'', x' ; \beta) \right).
\end{equation}    
\end{widetext}

 The ${G}^{SHO}(x, x''; \beta) $ can be computed from path integral kernel for SHO \cite{feynman2010quantum} by setting $T=\beta \hslash/i$ to obtain
  \begin{equation}
      {G}^{SHO}(x, x''; \beta)  = A \exp{-B \left[ (x^{2}+ x''^{2}) C - 2 x x''   \right]} . 
  \end{equation}
  Here $A= \sqrt{\frac{m \omega}{2 \pi  \hslash \sinh({\omega \beta \hslash})}}$, $B= \frac{m \omega}{2 \hslash \sinh({\omega \beta \hslash})}$ and $C=\cosh({ \omega \beta \hslash}) .$
 The ${G}^{0}(x, x''; \beta) $ is similarly obtained as
\begin{equation}
     {G}^{0}(x'', x'; \beta) = A_{0} \exp{-B_{0} (x''-x')^{2}}.
\end{equation}
Here, $A_{0}=\sqrt{\frac{m}{2\pi \beta \hslash ^{2}}} $, $B_{0}= \frac{m}{2 \beta \hslash^{2}}$. 

Following the steps elaborated in Appendix \ref{app1dsho} we finally obtain the total GF as 
\begin{widetext}
\begin{equation}\label{avggf}
\bar{G}(x,x';\beta)=G^{SHO}(x,x';\beta)+\frac{1}{2} \left( \left(\delta - \zeta \right) e^{-\alpha(ax^2+ex'^2 -bxx')} + \left(\delta - \zeta' \right)e^{-\alpha(ax'^2+ex^2 -bxx')}\right)
\end{equation}
\end{widetext}

Here
\begin{subequations}\label{paramsgf}
\begin{align}
\delta &= \frac{3 \left( e^{-\beta W_{0}}-1\right) }{2}A A_{0}\sqrt{\frac{\pi}{\phi}}, \\ 
\zeta(x,x') &=  A A_{0}B_{0}\left( e^{-\beta W_{0}}-1\right)\sqrt{\frac{\pi}{\phi}}\left[\frac{1}{2 \phi}+(\theta(x,x'))^2 \right],\\
\zeta'(x',x) &= A A_{0}B_{0}\left( e^{-\beta W_{0}}-1\right)\sqrt{\frac{\pi}{\phi}}\left[\frac{1}{2 \phi}+(\theta'(x',x))^2 \right],\\
\phi &= (BC+B_{0}),\\
\theta (x,x') &= \frac{B(x-Cx')}{\phi},\\
\theta' (x',x) &=  \frac{B(x'-Cx)}{\phi},\\
 a&=B^2C^2-B^2+BCB_{0},\\
    e&= BCB_0, \\
    b&=2B B_{0},\\
    \alpha &=1/\phi.
   \end{align} 
\end{subequations}

\begin{figure}
    \centering
    \includegraphics[scale = 0.4, angle=270]{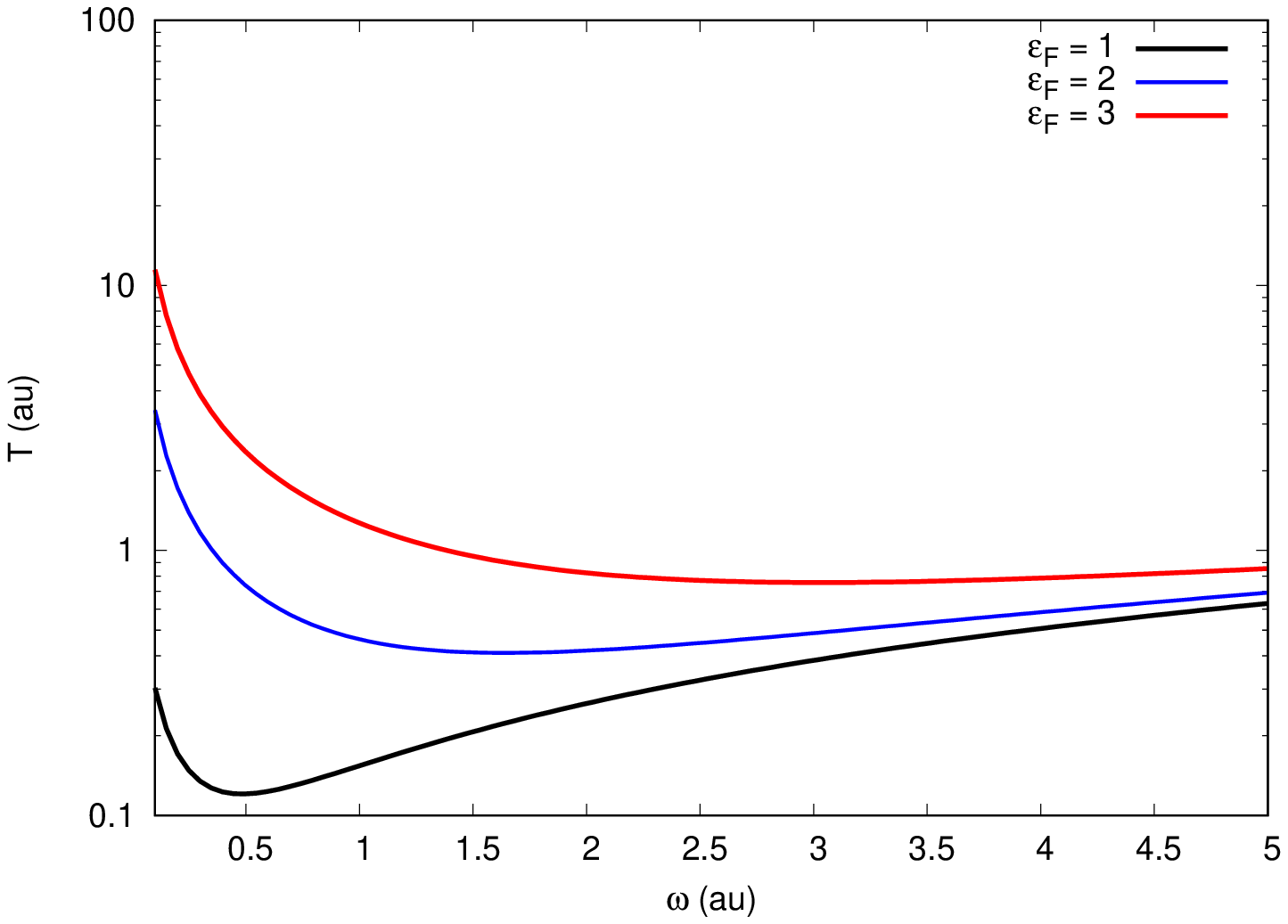}
    \caption{Variation of kinetic energy with $\omega$ for different Fermi energies $\ef$, keeping $m = 1$ and $W_0 = 1$. Note that the $\omega_{\text{min}}$ shifts to higher values with an increase in $\ef$.  The ordinate is in logscale.}
    \label{fig:t_vs_w}
\end{figure}

Following our formalism, we finally obtain the kinetic energy 
\begin{equation}\label{ketbeta}
    T[\rho] = \lim_{\Gamma \to \infty}\frac{1}{2\pi i} \int_{\gamma-i \Gamma}^{\gamma+i\Gamma}\frac{\text{d}\beta}{\beta}e^{\beta \ef} t(\beta)
\end{equation}
where $t(\beta)$ is written as
\begin{widetext}
\begin{equation}
    t(\beta)= \left(\frac{m}{8 \omega \hslash^3 \beta^2 }+ \frac{m \omega   }{8 \hslash} 
    +\frac{e^{-\beta W_{0}}-1 }{2} \times\left( \frac{9 m \omega }{32\sqrt{2} \hslash}+ \frac{3m}{4\sqrt{2}\omega \hslash^3 \beta^2} +\frac{m}{2\sqrt{2}\omega \hslash^3 \beta^2} +\frac{43 m \omega}{128\sqrt{2} \hslash} \right)\right]
\end{equation}    
\end{widetext}

leading to the final KE 
\begin{equation}\label{tot_ke}
    T[\rho] = \frac{m\omega}{8 \hslash}+\frac{m \epsilon_{F}^{2}}{4 \omega \hslash^3}-\frac{5m\epsilon_{F}^{2}}{16\sqrt{2}\omega\hslash^3}+\frac{5 m (\epsilon_{F}-W_{0})^2}{16\sqrt{2} \omega \hslash^3}
\end{equation}
Here $\epsilon_{F}$ is the Fermi energy. We can see clearly from Eq.\eqref{tot_ke} that the kinetic energy increases quadratically with $\ef = \frac{p^2_F}{2m}$ where $p_F$ is the Fermi momentum. $p_F$ signifies the highest possible momentum of the particles in the system. As a result, the involvement of higher momentum states leads to higher kinetic energy for the system. On the contrary, the repulsive interaction term $W_0$ pushes particles away from each other as well as from the center of the SHO. Classically, the momentum of the SHO is maximum at the center. This effect leads to a lower KE with the increase in $W_0$. The variation of the KE with $\omega$ is more complex. It is evident from Fig.\ref{fig:t_vs_w} that there exists a frequency $\omega_{min}$ at which the kinetic energy is minimum for a given $\ef$ and $W_0$. 
It is important to note here that $W_0 << \ef$ should be considered as the weak field limit (see Sec. \ref{weakfieldlim} for a more mathematically motivated rationale). Physically, this definition of weak-field limit can be justified as follows. The Fermi energy $\ef$ represents the highest energy states for the non-interacting Hamiltonian which is compensated by central potential provided via a positive force constant. In realistic atoms, this central potential on negatively charged electrons is exerted by the positively charged nucleus via Coulomb interaction. If the inter-particle repulsive interaction $W_0$ supersedes the Fermi energy, the system will become unstable and unbound. As a result, inter-particle interaction cannot be computed as a perturbative effect. 
We can also obtain the expression of $\omega_{min}$ analytically (see Appendix \ref{omwgamincalc}) as 
\begin{equation}\label{omegamin}
    \omega_{min} \approx \left(\frac{\ef}{\hslash}\right)\left(2-\frac{5W_0}{\sqrt{2}\ef}\right)^{1/2}
\end{equation}
In other words, the kinetic energy, and by virtue of the virial theorem the total energy, gets minimized in the presence of interelectronic interaction for a critical frequency $\omega_{min} \propto \sqrt{k_{min}}$ .  The existence of $\omega_{min}$ is reminiscent of Slater's shielding constants where the interelectronic interaction reduces the effective nuclear charges experienced by an electron in a single-particle state. Since the force constant $k$, and therefore $\omega$, takes the role of the nuclear charge in our present work the $\omega_{min}$ hints at the similar phenomenology. Also, as evident from Eq.\eqref{omegamin}
 this $\omega_{min}$ increases with the $\ef$. This effect is parallel to the fact that the shielding due to inner electrons is more significant than those in higher energy states since these higher energy electrons are localized farther away from the center of attraction leading to lesser shielding.  

\subsection{Application of traditional approach for 1D SHO }\label{tradgf_1dsho}
 A direct term-by-term comparison between our approach and the traditional Green's function defined in Eq.\eqref{deftradgf} is warranted here. Considering the same  approximation of 1D SHO for traditional GF, Eq. \ref{traditional_gf} can be written as,
\begin{equation}\label{traditional_gf}
    \tilde{G}(x,x';\tau) = \mathcal{G}^{H}(x,x';\tau)+ \int dx'' \mathcal{G}^{H}(x,x'';\tau)W(x'')\tilde{G}(x'',x',\tau).
\end{equation}
 The idea is to compare our GFF approach with the traditional GF approach using the same approximation, whereas explained above we replace $G^{H}(x,x';\beta)$ by $G^{SHO}(x,x';\beta)$ and $W(x'')$ is the mean-field potential as explained in Eq.\ref{weakfield}, here also considered the constant interaction potential.
 \begin{equation}\label{traditional_gf}
    \tilde{G}(x,x';\tau) = \mathcal{G}^{SHO}(x,x';\tau)+ \int dx'' \mathcal{G}^{SHO}(x,x'';\tau)W_{0}\tilde{G}(x'',x',\tau).
\end{equation}

The form of  $\mathcal{G}^{SHO}$ with energy $E$ is \cite{Chua_2018SHOGF} 
\begin{equation}\label{tradgfshofull}
    \mathcal{G}^{SHO}(x,x';E)= \frac{m \Gamma(\frac{1}{2}-\epsilon)}{2^{\epsilon -1/2} \sqrt{\pi} \alpha \hslash^2} e^{- \frac{\alpha^2 (x^2+x'^2)}{2}} H_{\epsilon -1/2}(-\alpha \min(x,x')) H_{\epsilon-1/2}(\alpha \max(x,x')).
\end{equation} 
Here $\alpha = \sqrt{\frac{m \omega}{\hslash}}$, $\epsilon = \frac{mE}{\alpha^2\hslash^2}= \frac{E}{\hslash \omega}$ and $H_{\epsilon -1/2}(x)$ is Hermite polynomial of degree $\epsilon-1/2$.  Note that to compute  $\mathcal{G}^{SHO}$, $E$ should be replaced by $\tau$. We decided to compute the kinetic energy from this expression for the future due to its complex form. Nevertheless, it is clear from  Eq.\eqref{tradgfshofull} that the leading order term of the interaction part of Green's function varies linearly with $W_0$. This is in contrast with our approach where the leading order interaction term (See Eq.\eqref{paramsgf}) is proportional to $\left(e^{-\beta W_0}-1\right)$ thereby incorporating $W_0$ contribution up to infinite order. Also, the $W_0$ appears in the exponent ensuring the convergence of the resulting series. Clearly, each term of our perturbation series appears to be a partial sum of infinite terms of the Dyson series presented in Eq.\eqref{dyson}. The exact nature of this partial sum, however, requires a sophisticated Feynman diagram approach which will be taken up in the future.       
\subsection{Spatial nature of Green's functions}\label{gfnature}

The fundamental entity in our formalism is the Green's function. Eq.\eqref{gf_shofree} is the governing equation for GF of a one-dimensional SHO in the weak interaction limit. In a broad sense, any Green's function captures the non-local response of the system, thereby encapsulating all mechanical information about the system. One can interpret the GF as the magnitude of the ``effect'' experienced by an observer at $x'$ when a ``cause'' has been introduced at $x$. Note that these ``cause-effect'' points are arbitrary and therefore the GF is symmetric with respect to $x$ and $x'$. Therefore it is imperative to analyze the spatial nature of the GF to understand and propel our methodology further. We have plotted Eq.\eqref{avggf} in Fig.\ref{fig:Nature_GF} for three representative $\omega = 0.5, 1 \text{ and } 2$ keeping  $\beta =10$ and $W_{0} = 1$ fixed. The contour plots in Fig.\ref{fig:Nature_GF} for $G^{SHO}$, $G^{int}$, and the total $G$ are generated using different parameter values. As expected from Eq.\eqref{avggf}, all contour plots exhibit $x-x'$ symmetry. Graphically, all contours are symmetrically distributed around the diagonal line for which $x=x'$.  The contour of $G^{SHO}$ is completely circular owing to the fact that the $G^{SHO}(x, x',;\beta)$ is a function of $(x-x')$. It also exhibits a maximum response at the center  ($x=x'=0$) due to the accumulation of probability density $\rho^{SHO}(x)$ at the center as described by Eq.\eqref{rhosho}. The response decreases as one moves away from the center since $\rho^{SHO}(x)$ also tapers off from the center monotonically. Similarly, $G^{int}$ shows the maximum response at the center albeit with a negative sign. As a result, the total Green's function represents a ``tug-of-war'' between the unperturbed SHO response and that of the interaction term. $G^{int}(x, x';\beta)$ also tapers off away from the center due to a decrease in probability density farther away. In contrast to $G^{SHO}$, $G^{int}$ shows a very prominent ``polar'' structure. One can see that with increase in $\omega$ the $G^{int}$ becomes more localized mirroring $\rho^{SHO}(x)$. Clearly, the response is longer in range along $x=0$ and $x'=0$ lines. This feature becomes more prominent with the increase in $\omega$. While the exact reason for this feature is not clear at present, we believe that it is related to the density difference between ``cause'' and ``effect'' points. Since $x=0$ or $x'=0$, line represents the region maximum probability of the particle, $G^{int}$ along these lines indicate a large difference of density leading to stronger interaction indicating a strong correlation between the density gradient $\frac{\partial \rho^{SHO}(x)}{\partial x}$ and interparticle interaction $W_0$. This feature is clearly reminiscent of inhomogeneity in atomic electron density exemplified by atomic shell structures. Clearly, this interaction term appears to hold the key to atomic shell structures. However, more work is necessary to prove this conclusion explicitly. The total $G$ is a combined effect of $G^{SHO}$ and $G^{int}$ and is more difficult to understand physically. It is clear, however, that while the $G^{int}$ becomes less ``polar'' with the decrease in $\omega$, total GF $G(x, x';\beta)$ becomes more ``polar'' indicating a complex interplay between $G^{SHO}$ and $G^{int}$. Future works are essential to understand these features for devising novel methodologies applicable to realistic systems. 


\newcommand{\gshotwo}{\includegraphics[angle= 270,width=8em ]{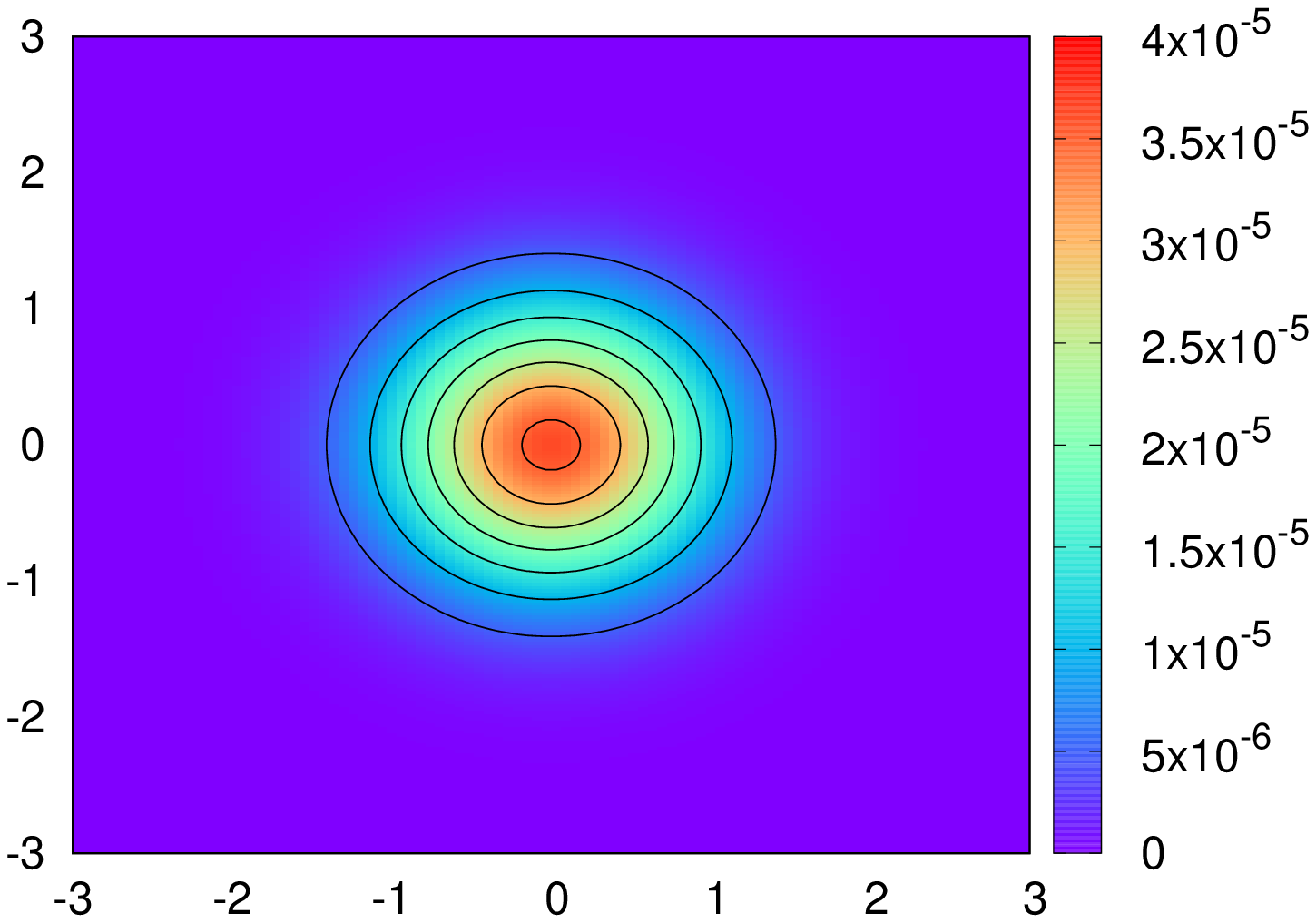}}
\newcommand{\ginttwo}{\includegraphics[angle= 270,width=8em ]{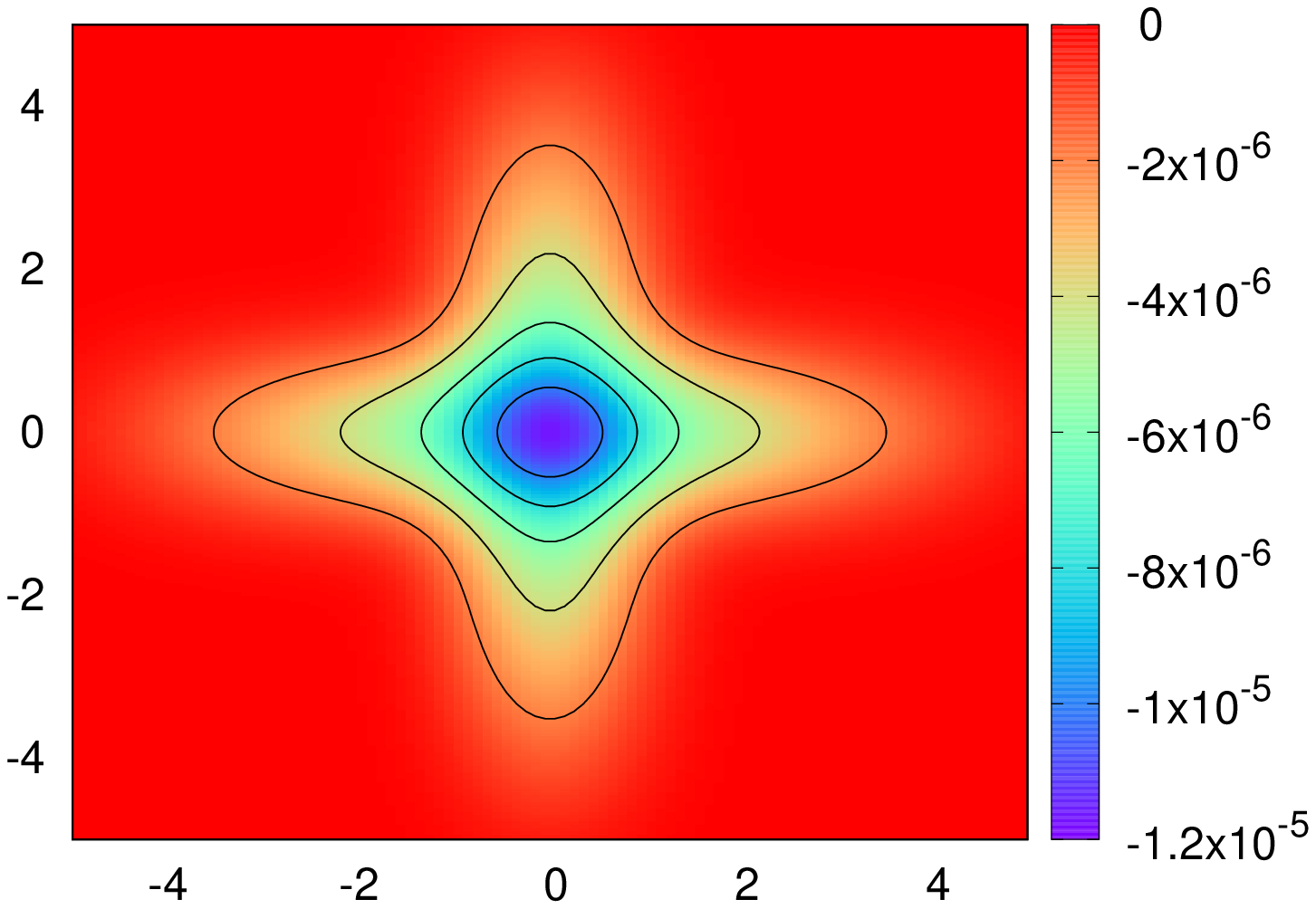}}
\newcommand{\gtwo}{\includegraphics[angle= 270,width=8em ]{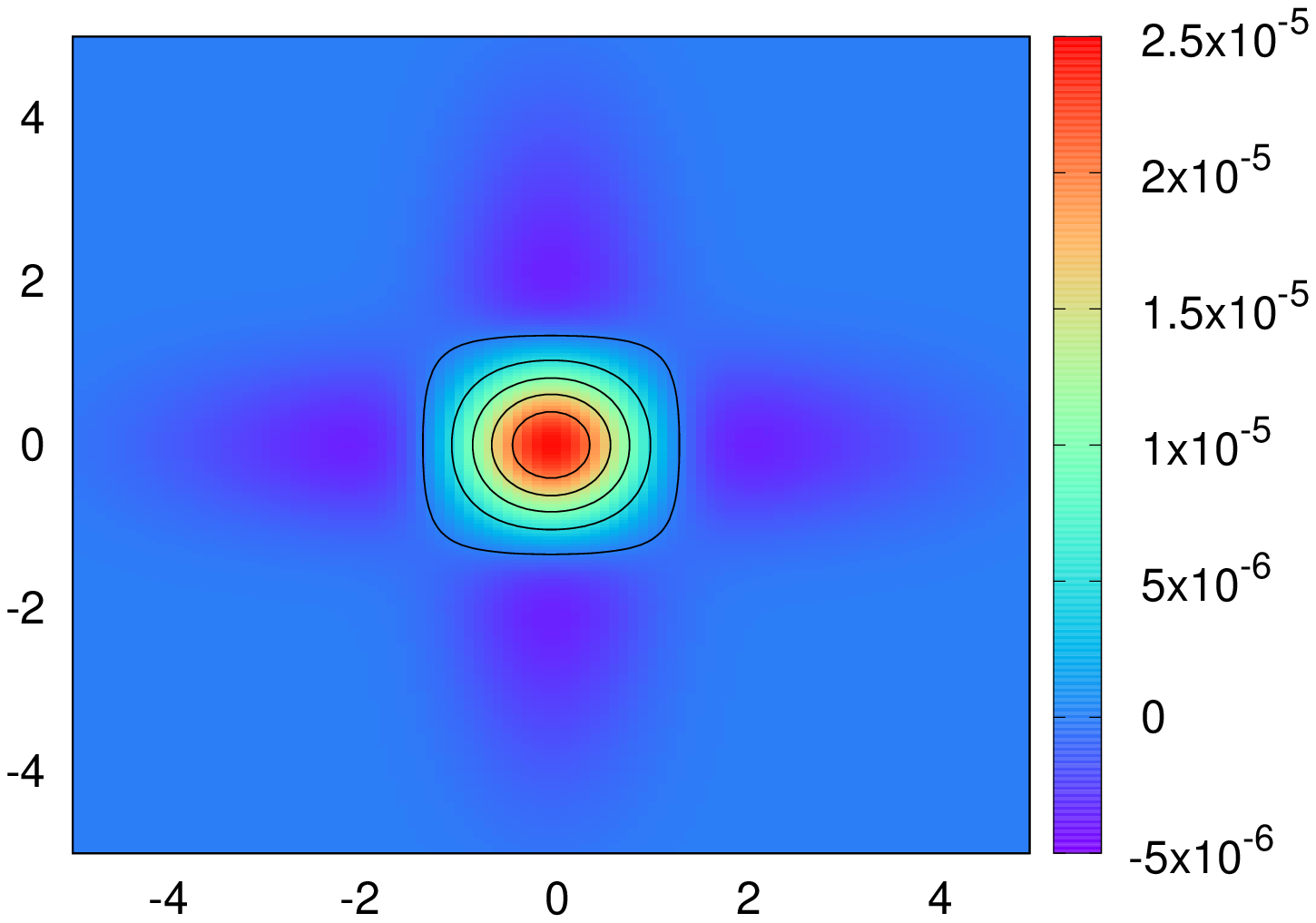}}
\newcommand{\gshoone}{\includegraphics[angle= 270,width=8em ]{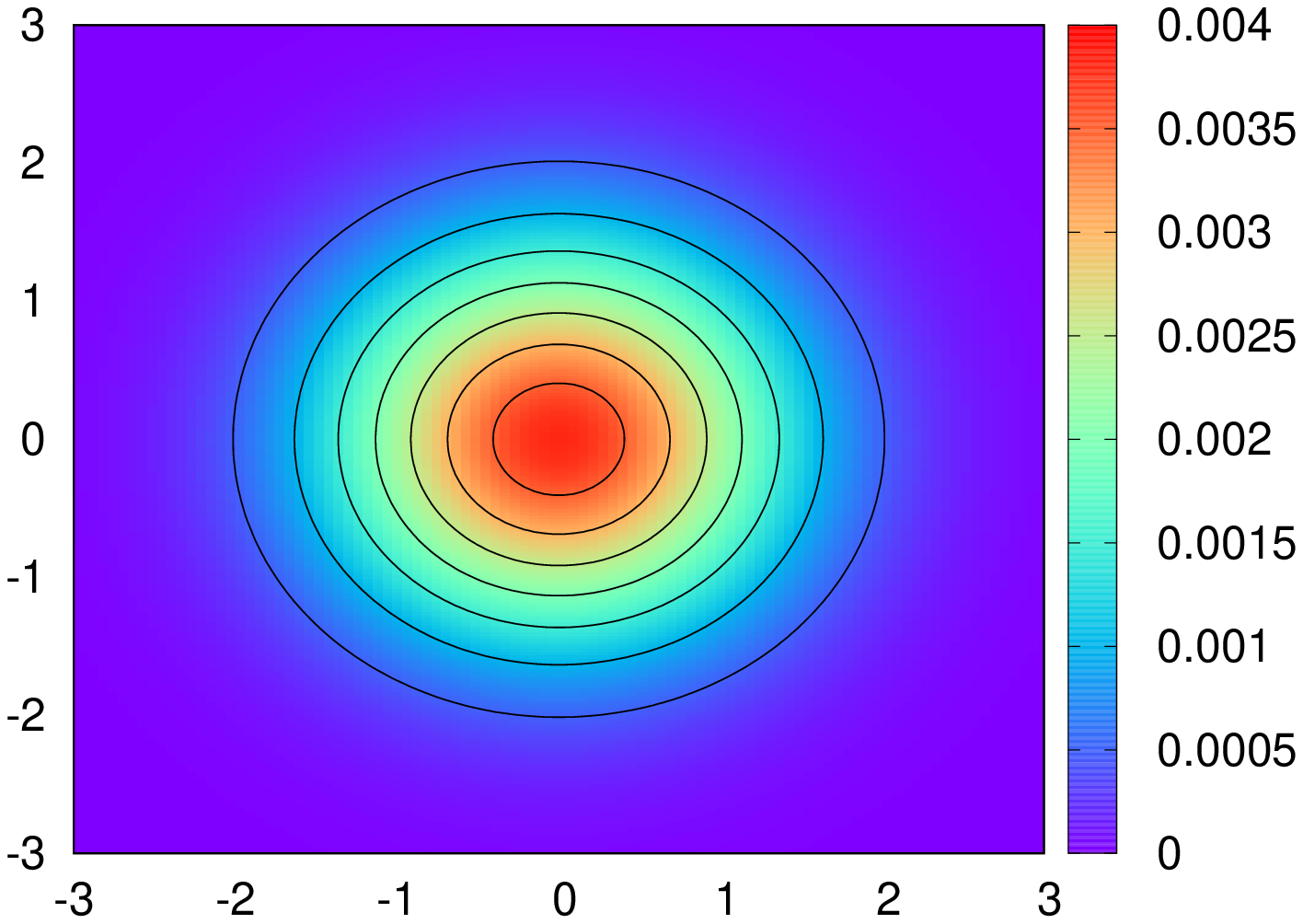}}
\newcommand{\gintone}{\includegraphics[angle= 270,width=8em ]{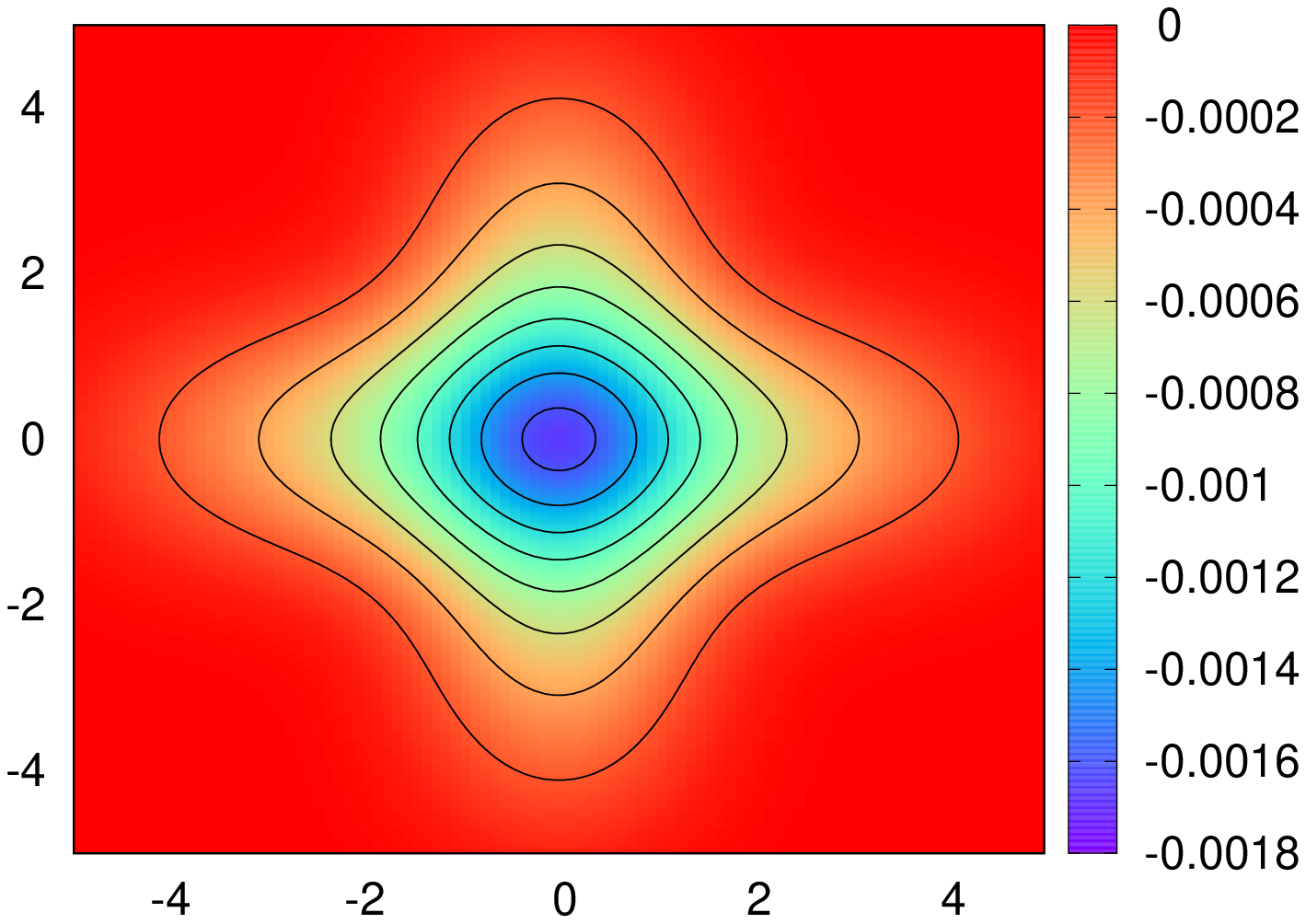}}
\newcommand{\gone}{\includegraphics[angle= 270,width=8em ]{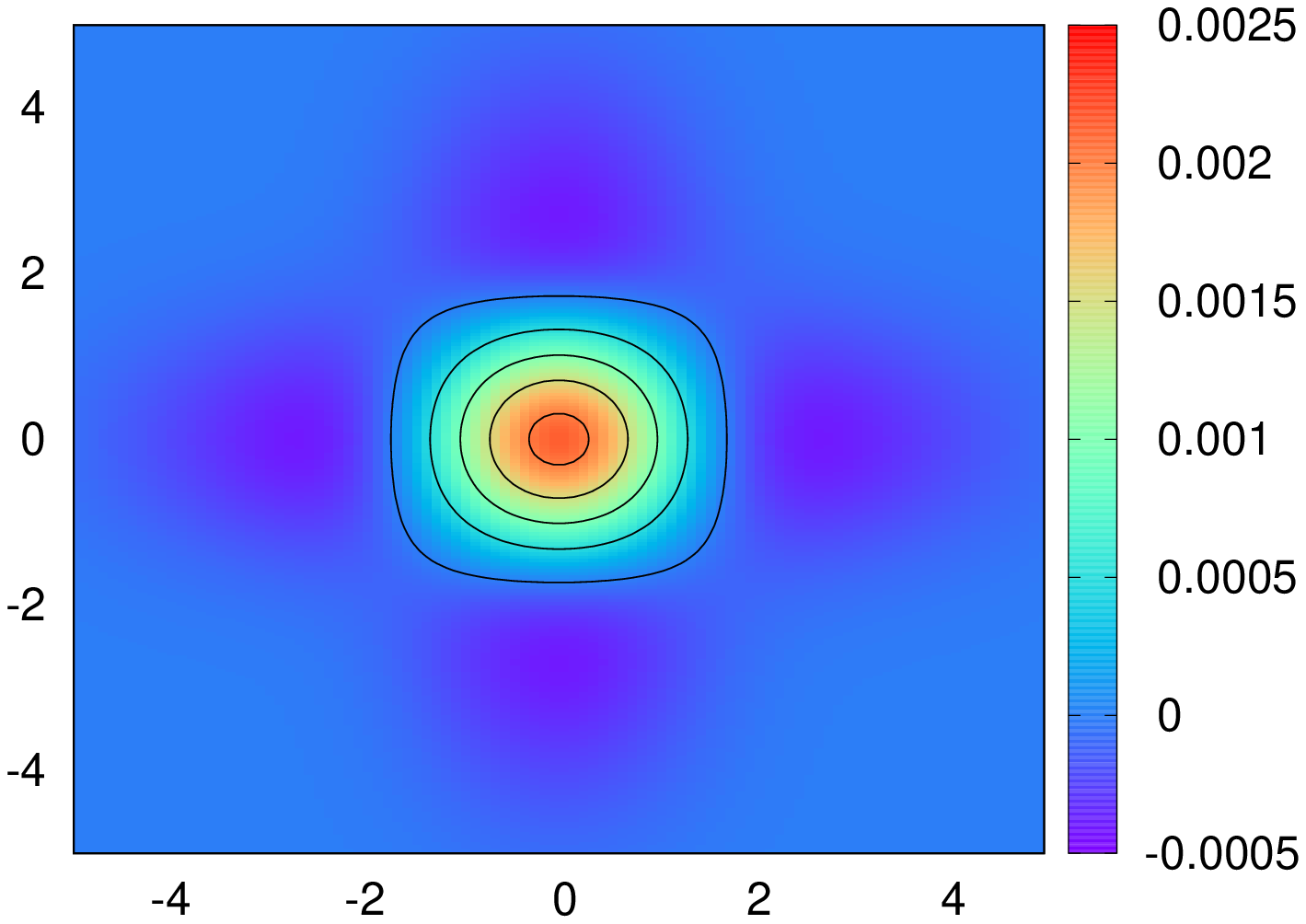}}
\newcommand{\gshohalf}{\includegraphics[angle= 270,width=8em ]{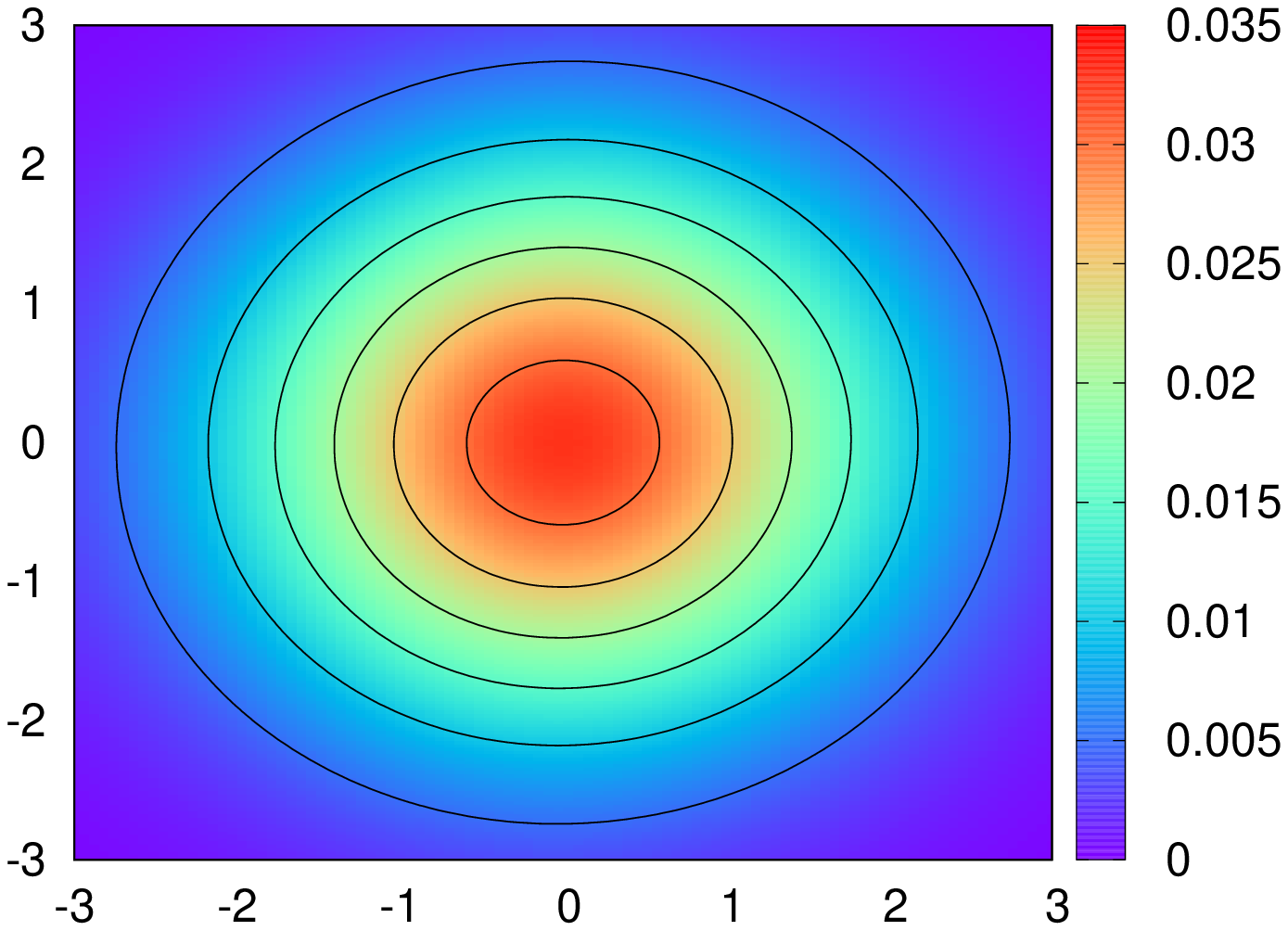}}
\newcommand{\ginthalf}{\includegraphics[angle= 270,width=8em ]{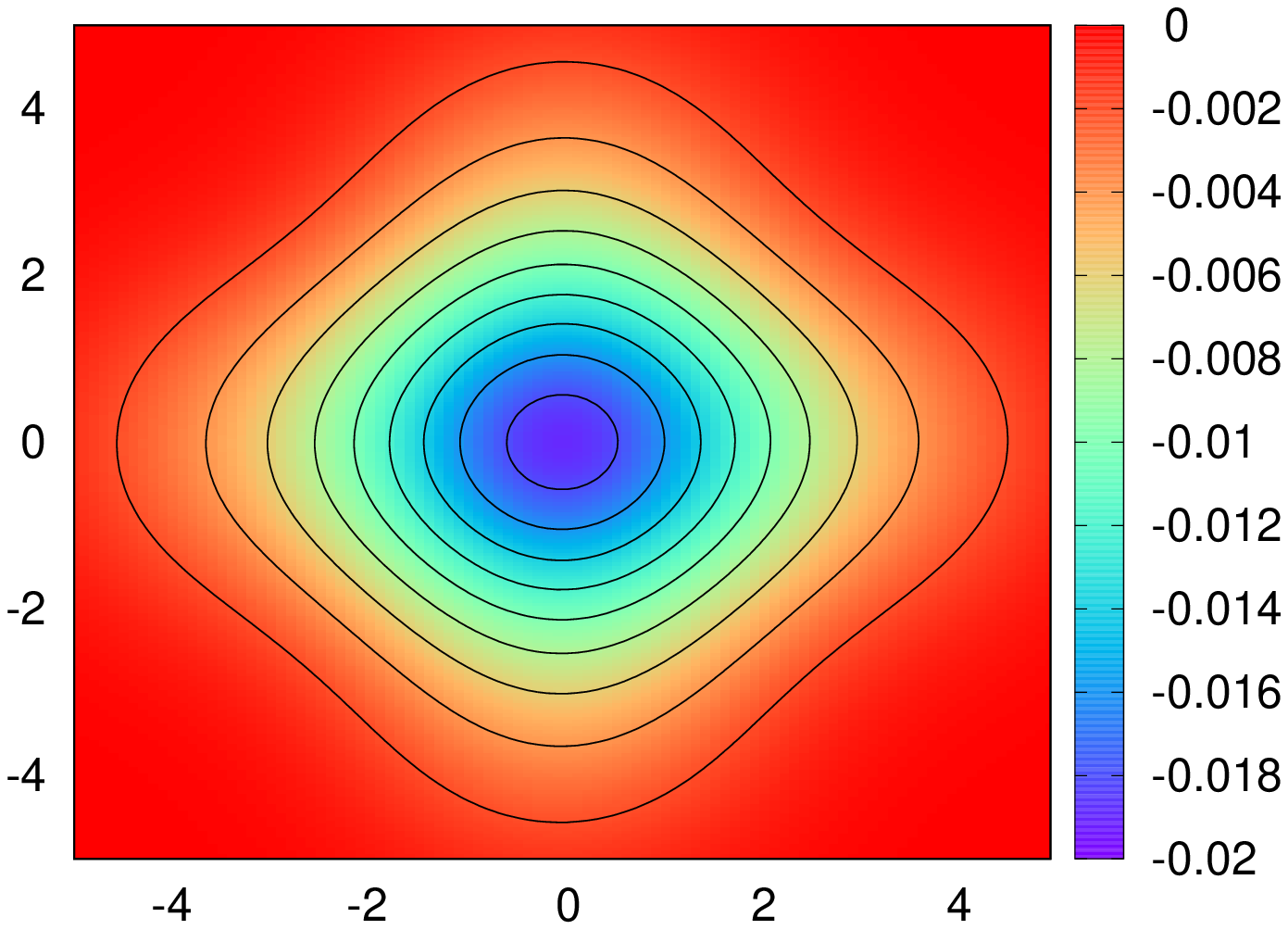}}
\newcommand{\ghalf}{\includegraphics[angle= 270,width=8em ]{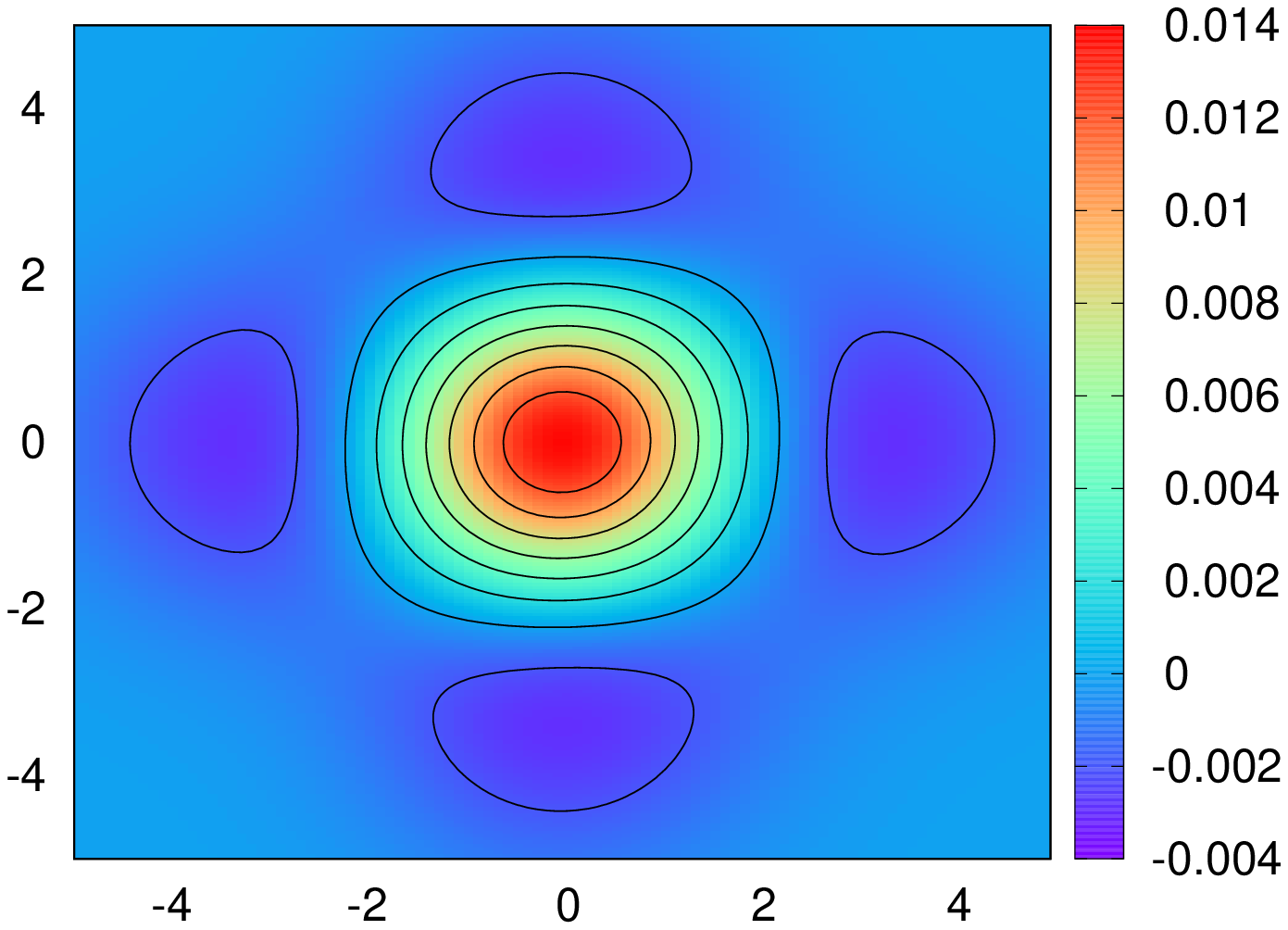}}


\begin{figure}[H]
\begin{tabular}{p{.2\textwidth}p{.2\textwidth}p{.2\textwidth}p{.2\textwidth}}
& \hspace{1cm}\textbf{A} & \hspace{1cm}\textbf{B} & \hspace{1cm}\textbf{C}\\
\vspace{0.5cm} \hspace{1cm}$\mathbf{\omega =2}$
&\gshotwo
&\ginttwo
&\gtwo\\
\vspace{0.5cm} \hspace{1cm}$\mathbf{\omega =1}$ 
&\gshoone
&\gintone
&\gone \\
\vspace{0.5cm} \hspace{1cm}$\mathbf{\omega =0.5}$
&\gshohalf
&\ginthalf
&\ghalf\\
\end{tabular}
\caption{The contour plots of $G^{\text{SHO}}$, $G^{\text{int}}$, and $G$ for different $\omega$ values and a constant $\beta = 10$. The first column (A) displays $G^{\text{SHO}}$, the second column (B) shows $G^{\text{int}}$, and the third column (C) illustrates $G$. The rows correspond to different $\omega$ values: the first row is for $\omega = 2$, the second row is for $\omega = 1$, and the third row is for $\omega = 0.5$. Each contour plot provides a spatial nature of the corresponding GFs under the specified parameters.
}
\label{fig:Nature_GF}
\end{figure}

\subsection{$N$ dependence of KE}\label{nzdep}
   Next, we will discuss the variation of the KE expression obtained in Eq.\eqref{tot_ke} with the number of particle $N$ via $\ef$ and $\omega$. We realize that $\ef$ indicates the upper limit of the occupied energy states and is therefore indicative of $N$ in a given set of single-particle states. We have employed two estimates of $\ef$ in this work resulting in two asymptotic limits. They represent pure bound state limit (BSL) and pure scattering state limit (SSL). This is done keeping in mind that realistic systems like hydrogenic atoms sustain both types of states and therefore these two models act as as asymptotic limits of the real atoms. Also, we considered each single-particle state to be singly occupied to mimic the effects of the Pauli exclusion principle. First, we consider the quantum harmonic oscillator as a representative of BSL to define $\ef$. As a result, we define bound state limit (BSL) by $\epsilon_F = (N+1/2) \hslash \omega$.  The kinetic energy functional for BSL is therefore
\begin{equation}\label{tbsl}
    T_{BSL} \approx A (N+1/2)^{2}- B(N+1/2)+C+D
\end{equation}
where $m$ and $\omega$ dependent constants A , B, C and D are 
\begin{eqnarray}
    A &=& \frac{ m \omega}{4 \hslash}\\
    B &=& \frac{5 m W_{0} }{8 \sqrt{2}\hslash^2}\\
    C &=& \frac{m \omega}{8 \hslash}\\
    D &=& \frac{5 m W_{0}^2}{16 \sqrt{2} \omega \hslash^3}.
\end{eqnarray}
 The SSL is represented by free particles with a continuous energy spectrum. In this limit 
 \begin{equation}
     \epsilon_{F} = \frac{\hslash^2}{2m} \left(\frac{3 \pi^2}{V} \right)^{2/3} N^{2/3}
 \end{equation}
 where the volume $V$ is computed here via classical turning points $x_{tp}$ of the SHO \textit{i.e} $V= (2x_{tp})^3$.
We employed the classical turning points are related to the energy of the highest occupied state $E_{N}= \left( N+ \frac{1}{2}\right) \hslash \omega$ and given by 
 \begin{equation}
     x_{tp}= \sqrt{\frac{2 E_{N}}{m \omega^2}}
 \end{equation}
 As a result, we obtain 
  \begin{equation}
      V=8 \left(\frac{2 (N+ 1/2) \hslash}{m \omega}\right)^{3/2}
  \end{equation}
  leading to the $\ef$
  \begin{equation}
      \ef = \frac{3\pi^2\hslash \omega  }{16} \frac{ N^{2/3}}{(N+1/2)}.
  \end{equation}
  Employing these expressions we finally find
\begin{equation}\label{tssl}
    T_{SSL} \approx E \frac{N^{4/3}}{(N+1/2)^2} -  F  \frac{N^{2/3}}{(N+1/2)}+ C+ D.
\end{equation}
where
\begin{eqnarray}
    E &=& \left(\frac{3\pi^2  }{16} \right)^2 \left(\frac{ m \omega}{4  \hslash} \right)\\
    F &=& \left(\frac{3\pi^2 }{16} \right) \left(\frac{5 m W_{0}}{8 \sqrt{2}  \hslash^2} \right)
\end{eqnarray}
To compare the relative qualitative accuracy of the four aforementioned KE functionals, we employed empirical $N$ and $Z$-dependency of kinetic energy $T[\rho]= T_{vW}[\rho]+ \gamma(N, Z) T_{0}[\rho]$ defined for atomic systems. For neutral atoms $N=Z$ leads to $\gamma(N,N)= 1-\frac{1.412}{N^{1/3}}$\cite{ParrYang1994,Acharya_1980}. Furthermore, to get the $N-$dependence of KE we employ $\rho \propto N$ to finally obtain 
\begin{equation}\label{tg}
        T_{\gamma} \approx N^{2}+ N^{5/3}- (1.412) N^{4/3}.
\end{equation}

\begin{figure}[H]
     \centering
         \includegraphics[  scale=0.5, angle =270 ]{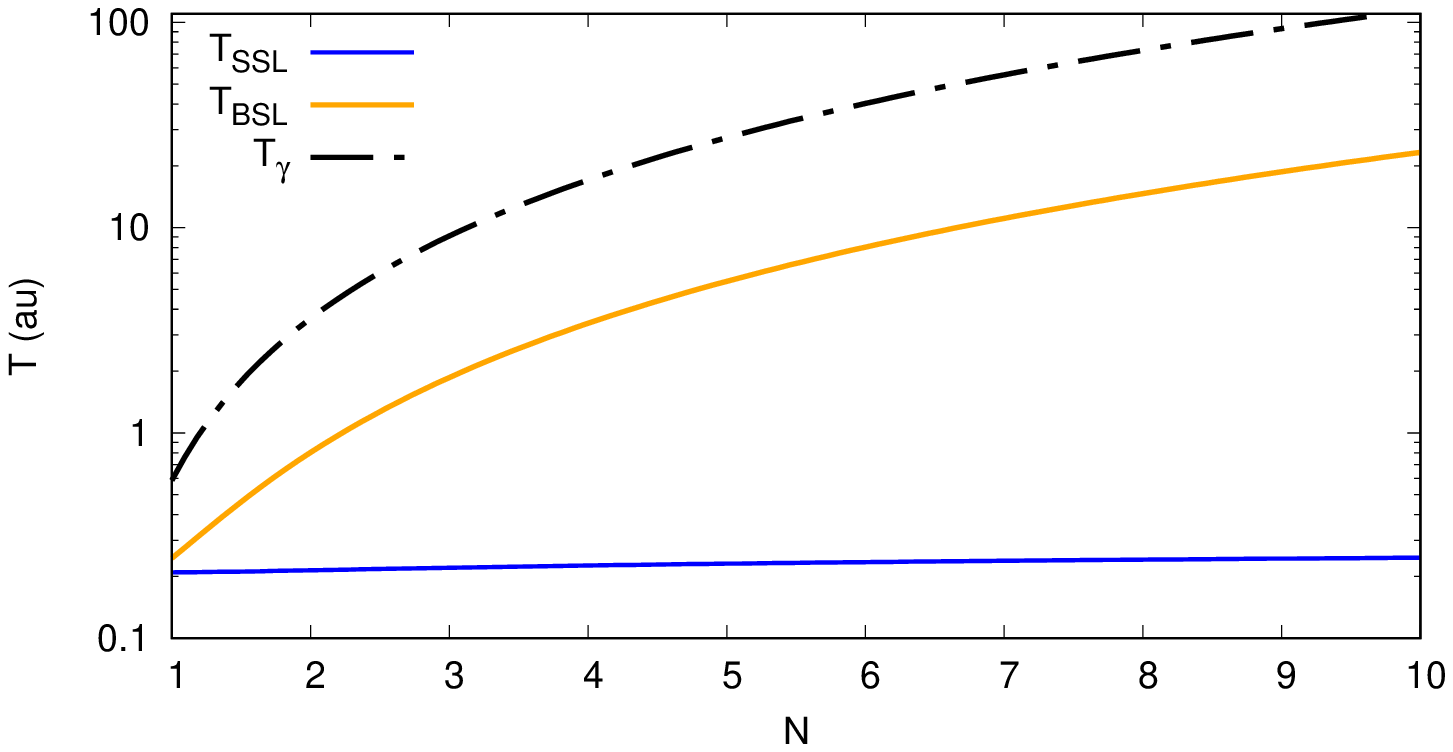} 
         \caption{The variation of kinetic energy as a function of $N$. The blue solid line, yellow solid line, and black dashed line represent $T_{SSL}$, $T_{BSL}$  and  $T_{\gamma}$, respectively. The ordinate is in logscale.}
         \label{fig:T_N}
         \end{figure}
A direct comparison between $T_{BSL}$, $T_{SSL}$ and $T_{\gamma}$ in Fig.\ref{fig:T_N}. It is evident that the $T_{BSL}$ follows the same functional dependence on $N$ as $T_{\gamma}$. The behavior of $T_{SSL}$ is completely wrong. This is expected since the realistic atoms are closer in essence to BSL than SSL. The difference between $T_{\gamma}$ and $T_{BSL}$ also increases for larger $N$. This effect is also expected since the single-particle states in SHO remain equispaced in energy while for atoms they become more closely spaced. Therefore for larger $N$ the qualitative difference between our model and actual atoms also gets magnified leading to larger error. While a quantitative agreement between our model and an actual atom is completely unexpected, our model appears to reproduce the qualitative features of realistic atoms, including the $N$-dependence of the KE, quite successfully.      
\subsection{Application of $T_{BSL}$ on noble gas atoms}\label{noblegas}
To push our formalism further, we have now applied our results to noble gas atoms. Note that it is expected that our results will not match the actual kinetic energy numerically for our expression is derived for a one-dimensional (1D) system. Also, it incorporates other crude approximations. However, we are interested in finding out if there is any qualitative trend that emerges from $(T_{BSL})$ applied to noble gas atoms using Eq.\eqref{tbsl}. The reference kinetic energy values were taken from Hartree-Fock table data \cite{Clementi1974}. We employed two estimates for $\omega$. First, we have used the ionization energies of the corresponding atoms in atomic units (\textit{a.u.}) as an estimate of $\omega= \omega_{IE}$ \cite{NIST_ASD}. The second estimate of $\omega = \omega_{K-L}$ is taken to be the transition frequency between K and L shells of the respected atoms. Results from these two estimates are presented in Table.\ref{tab:ke_table}. Note that except for He where these are essentially the same due to following Koopmans theorem, for other atoms $\omega_{IE} < \omega_{K-L}$. Moreover $\omega_{IE}$ and $\omega_{K-L}$ represent the smallest and largest energy gaps for these atoms. Therefore it is expected that actual values of the KE will lie between these two extreme limits. This expectation is indeed found to be true. It can be seen from Table \ref{tab:ke_table} that $T_{BSL}$ and $T_{BSL}^{K-L}$ represent lower and higher extremes of true KE. In the future, a search for an effective frequency may be pursued which will reproduce the KE more faithfully. Such quantity will turn out to be very useful for Unsold-type approximations in the computations of effective response functions \cite{tim_fred_unsold_2018}.  
\begin{table}[h]
    \centering
\begin{tabular}{|c|c|c|c|c|c|}
\hline
\textbf{Atom}  & $\mathbf{\omega_{IE}} $ (\textit{a.u.})   &  $\mathbf{T_{BSL}}$ (\textit{a.u.}) & $\mathbf{\omega_{K-L}}$ (\textit{a.u.}) & $\mathbf{T_{BSL}^{K-L}}$ (\textit{a.u.}) & \textbf{K.E} (\textit{a.u.}) \\ [0.5ex] 
  \hline
\textbf{He} & $9.1 \times 10^{-1}$ & $6.7 \times 10^{-1}$ & $9.2 \times 10^{-1}$ & $6.8 \times 10^{-1}$ & $2.86 \times 10^{0}$ \\
   \hline
\textbf{Ne} & $7.9 \times 10^{-1}$ & $1.76 \times 10$ & $3.08 \times 10^{0}$ & $8.49 \times 10^{2}$ & $1.28 \times 10^{2}$ \\
 \hline
\textbf{Ar} & $5.8 \times 10^{-1}$ & $4.18 \times 10$ & $1.06 \times 10^{2}$ & $9.09 \times 10^{3}$ & $5.26 \times 10^{2}$ \\
 \hline
\textbf{Kr} & $5.1 \times 10^{-1}$ & $1.55 \times 10^{2}$ & $4.50 \times 10^{2}$ & $1.5 \times 10^{5}$ & $2.75 \times 10^{3}$ \\[1ex]
\hline
\end{tabular}
\caption{Calculated kinetic energy values, $T_{BSL}$ and $T_{BSL}^{K-L}$, for noble gas atoms using $\omega_{IE}$ and $\omega_{K-L}$, respectively. The reference kinetic energy (K.E.) values are taken from Hartree-Fock table data \cite{Clementi1974}. The parameters $\omega_{IE} $\cite{NIST_ASD} and $\omega_{K-L}$ are used to compute the $T_{BSL}$ and $T_{BSL}^{K-L}$ values, respectively.}
\label{tab:ke_table}
\end{table}

\subsection{A natural limit of weak-field interaction}\label{weakfieldlim}
We have explored the weak field limit  extensively in our work and defined it as 
\begin{equation}\label{wkfielddef}
W_0^{max} < \ef    
\end{equation}
 in Section \ref{gfweakfield} from a physical consideration where $W_0^{max}$ is the maximum allowed value for interaction potential. In addition, our formalism supports this definition in a more mathematically rigorous manner. The condition to obtain a well-defined Laplace transform in Eq.\eqref{ketbeta} is 
\begin{equation}
    \ef - W_{0} \geq 0. 
\end{equation}
Without the condition, the Bromwich integral does not yield a bounded KE justifying our physical definition in Eq.\eqref{wkfielddef}. 
Note that $\ef$ increases with $N$. Therefore for $N\to \infty$, any system would be considered as the weak-field limit leading to an exact HEG description. This conclusion coincides with Lieb's results\cite{Lieb_RevModPhys.48.553}.

\section{Conclusion} \label{conclude}
In summary, the salient features of our present works are

    (i) the  development of a  general and systematic analytical method to compute the kinetic energy density for an electronic system using Green's function formalism, (ii) a rigorous mathematical connection of our present method to traditional Green's function method, (iii) indication of a clear source of $N$-dependence of KE and (iv) a natural bound of weak interaction strength between electrons. 
    
    The methodology developed in this work is the first approach, to our knowledge, to compute KED in a completely systematic manner \textit{without relying upon any model potential as the starting point}. Instead, the guiding principle of our perturbative treatment is the magnitude of the different partitions of the total Hamiltonian leading to a mathematically consistent methodology for KED. 
    We have also shown that our methodology is related to traditional Green's function formalism in a mathematically rigorous manner. This feature will help us to compare and borrow from the already developed series expansion techniques used in GW-based methods. 
    Each term of our expansion can be considered as a partial sum of the traditional perturbation series of Green's functions. To exemplify the formalism we also presented some flagship calculations on a one-dimensional Hookian atom. We find the KE values computed from this model provides a natural upper and lower bounds for KE of noble gas atoms depending on the excitation frequency used. These calculations successfully reproduce several qualitative features of atomic KEDs such as the the variation of KE on number of particles $N$. To the best of our knowledge, this is the first work which shows such dependence \textit{without recoursing to any asymptotic model}. 
    As a byproduct of our formalism, we also obtained an upper limit of the magnitude of interaction potential in terms of the Fermi energy as  a quantitative measure for ``weak'' potential. Furthermore, the mathematically rigorous connection between traditional Green's function and our Green's function allows for an exact recipe for the systematic development of KE, perhaps facilitating the use of diagrammatic expansion techniques.

Despite being exact, there will be a few directions which will be considered in near future to make this method computationally applicable for real atoms. First, the form of self-energy computed in this work is done in a mean-field framework. A more rigorous many-body description will be developed in the future. As a consequence, several sophisticated approximation techniques have to be developed. One of the next goals is to find $Z$-dependence of KE for atomic systems. In that pursuit, the use of Hydrogenic Green's functions appears to be a good starting point. Moreover, a diagrammatic perturbation technique is essential to develop an analytically convergent perturbation theory.

\begin{acknowledgments}
    Priya acknowledges a Prime Minister's Research Fellowship and MS acknowledges IIT Kanpur initiation grant no. IITK/CHM/2018419 and SERB startup research grant no.
SRG/2019/000369 for partial supports.  We also thank Jaseela for helping us to prepare the manuscript.  
\end{acknowledgments}

\begin{appendix}

\section{Connection to $G(\br, \br'; \beta)$ and $\tilde{G}(\br, \br';\tau)$}\label{gtogtilde}
We define the Fourier transform operator $\hat{\mathcal{F}_{\mu}^{\omega}}$ as 
\begin{equation}\label{FTdef}
    F(\omega)=\hat{\mathcal{F}_{\mu}^{\omega}}(f(\mu))\equiv \frac{1}{\sqrt{2\pi}}\int e^{i\mu \omega}f(\mu) \text{d}\mu.
\end{equation}
Using Eq.\eqref{FTdef}, we can our Green's function in terms of traditional GF as

\begin{eqnarray}
    \gfrrp &=& \lim_{\mu \to 1^+} \bra{\br}\hat{\mathcal{F}}_{\mu}^{\omega}(\frac{1}{\beta \ham+i\omega})\ket{\br'} \nonumber \\
    &=& \lim_{\mu \to 1^+} \frac{1}{\beta}\bra{\br}\hat{\mathcal{F}}_{\mu}^{\omega}(\frac{1}{ \ham+i\frac{\omega}{\beta}})\ket{\br'}\nonumber \\
    &=& \lim_{\mu \to 1^+} \frac{1}{\beta} \hat{\mathcal{F}}_{\mu}^{\omega}( \tilde{G}(\br, \br'; \tau)).
\end{eqnarray}
where $\tau = \frac{-i\omega}{\beta}$. 
\section{Calculation for 1D harmonic oscillator using GFF}\label{app1dsho}
The detailed calculation for kinetic energy is described here.
\begin{equation} 
   \gfxxp = {G}^{SHO}(x, x'; \beta)+ \int dx'' {G}^{SHO}(x, x''; \beta) \bra{x''}\sum_{n=0}^{\infty} \frac{\beta^n}{n!}\ham_{0}^{n}\left(e^{-\beta (\ham_0+\hat W)} - e^{-\beta \ham_{0}}\right) \ket{x'}.  
\end{equation}
Basic definition used in our work 
\begin{equation}
      {G}^{SHO}(x, x''; \beta)  = A \exp{-B \left[ (x^{2}+ x''^{2}) C - 2 x x''   \right]} . 
  \end{equation}
  
Here
\begin{eqnarray*}
  A&=& \sqrt{\frac{m \omega}{2 \pi  \hslash \sinh({\omega \beta \hslash})}}\\
  B&=&\frac{m \omega}{2 \hslash \sinh({\omega \beta \hslash})} \\
  C&=&\cosh({ \omega \beta \hslash})
\end{eqnarray*}
 The ${G}^{0}(x, x''; \beta) $ is similarly obtained as

\begin{equation}
     {G}^{0}(x'', x'; \beta) = A_{0} \exp{-B_{0} (x''-x')^{2}}.
\end{equation}
\begin{eqnarray*}
   A_{0}&=&\sqrt{\frac{m}{2\pi \beta \hslash ^{2}}} \\
    B_{0}&=& \frac{m}{2 \beta \hslash^{2}}. 
\end{eqnarray*}
as explained above in section \ref{results}, we considered weak interaction limit 
\begin{equation}
     W(x) = W_{0} + x \partial_{x} W(x)|_{0}+ \hdots 
 \end{equation}
 We have considered the constant interaction potential, as a result
 eq. \eqref{GFmain} becomes
 \begin{equation}
     \gfxxp = {G}^{SHO}(x, x'; \beta)+ \int dx'' {G}^{SHO}(x, x''; \beta) \bra{x''}\sum_{n=0}^{\infty} \frac{\beta^n}{n!}\ham_{0}^{n}\left(e^{-\beta \ham_0}e^{-\beta \hat W_{0}} - e^{-\beta \ham_{0}}\right) \ket{x'}.  
 \end{equation}
 rewriting the above equation
 \begin{equation}
     \gfxxp = {G}^{SHO}(x, x'; \beta)+ \int dx'' {G}^{SHO}(x, x''; \beta) (e^{-\beta W_{0}}-1)\bra{x''}\sum_{n=0}^{\infty} \frac{\beta^n}{n!}\underbrace{ \ham_{0}^{n} e^{-\beta \ham_{0}}}_{\frac{\partial^{n}}{\partial(-\beta)^{n}} e^{-\beta \ham_{0}}} \ket{x'}. 
\end{equation}

\begin{equation}
  \gfxxp = {G}^{SHO}(x, x'; \beta)+ (e^{-\beta W_{0}}-1)\int dx'' {G}^{SHO}(x, x''; \beta) e^{-\beta \frac{\partial}{\partial\beta}} G^{0}  (x'',x'; \beta)
\end{equation}

solving for the 2nd part $e^{-\beta \frac{\partial}{\partial\beta} }G^{0}  (x'',x'; \beta) \approx \left (1-\beta \frac{\partial}{\partial \beta} \right)G^{0}  (x'',x'; \beta)= \frac{3}{2}G^{0}  (x'',x'; \beta)- B_{0} (x''-x')^2 G^{0}  (x'',x'; \beta)$

Therefore we get 
\begin{equation}\label{main1D}
\begin{split}
  \gfxxp = {G}^{SHO}(x, x'; \beta)+ (e^{-\beta W_{0}}-1) \left(\frac{3}{2}\int dx''{G}^{SHO}(x, x''; \beta)G^{0}  (x'',x'; \beta)\right)\\
  - (e^{-\beta W_{0}}-1)\left(B_{0}\int dx''{G}^{SHO}(x, x''; \beta) (x''-x')^2 G^{0}  (x'',x'; \beta) \right)  
\end{split}
\end{equation}
\section{The kinetic energy contribution from $G^{SHO}$ } \label{GSHO_cal}
The kinetic energy contribution from  $G^{SHO}$ in Eq.\eqref{main1D} can be calculated as
\begin{eqnarray}
    T[\rho]&=& -\frac{1}{2} \int {\frac{\partial^2}{\partial x^2} (\dmxxp)}\vert_{x=x'} dx
\end{eqnarray}    
As explained in section \ref{Theory} $\dmxxp$ is related to GF by Laplace transform
\begin{eqnarray}
    T[\rho]&=& -\frac{1}{2} \int {\frac{\partial^2}{\partial x^2}} \left( \lim_{T\to \infty}\frac{1}{2\pi i} \int_{\gamma-iT}^{\gamma+iT}\frac{\text{d}\beta}{\beta}e^{\beta \ef} \gfxxp \right) \vert_{x=x'} dx\\
    &=& -\frac{1}{2} \lim_{T\to \infty}\frac{1}{2\pi i} \int_{\gamma-iT}^{\gamma+iT}\frac{\text{d}\beta}{\beta}e^{\beta \ef} \int {\frac{\partial^2}{\partial x^2}} \gfxxp \vert_{x=x'} dx 
\end{eqnarray}
Here we are doing for $G^{SHO}$, therefore we need to evaluate ${\frac{\partial^2}{\partial x^2}}G^{SHO}(x,x';\beta)$
\begin{equation}
    \frac{\partial^2}{\partial x^2}G^{SHO}(x,x';\beta)\vert_{x=x'} = 4AB^2 x^2 (C-1)^2 e^{-2Bx^2(C-1)}- 2ABCe^{-2Bx^2(C-1)}
\end{equation}

\begin{equation}
    \frac{\partial^2}{\partial x'^2}G^{SHO}(x,x';\beta)= \frac{\partial^2}{\partial x^2}G^{SHO}(x,x';\beta)
\end{equation}
Now doing the integration w.r.t x
\begin{eqnarray}
\int  \frac{\partial^2}{\partial x^2}G^{SHO}(x,x';\beta)\vert_{x=x'} dx &=& \int dx (4AB^2 x^2 (C-1)^2 e^{-2Bx^2(C-1)}- 2ABCe^{-2Bx^2(C-1)}) \\
 & =& 4AB^2 (C-1)^2\frac{1}{2} \sqrt{\frac{\pi}{(2B(C-1))^3)}}-2ABC \sqrt{\frac{{\pi}}{{2B(C-1)}}}\\
 &=& A\sqrt{B\pi}\left(\sqrt{\frac{C-1}{2}}- \sqrt{\frac{2}{C-1}}C \right)\\
 &=&- A\sqrt{B\pi}\frac{C+1}{\sqrt{2(C-1)}}
\end{eqnarray} 
Now using $A=\sqrt{B/\pi}$ ,  $C=1+\frac{(\omega\beta\hslash)^2}{2}$ and using approximation $sinh(\omega\beta\hslash)=\omega\beta\hslash$, therefore  $B= \frac{m}{2\beta\hslash^2}$.
Using all the approximations we get
\begin{eqnarray}
    &=& -\frac{m}{4\beta\hslash^2} \left( \frac{4}{\omega\beta\hslash}+\omega\beta\hslash \right)\\
    &=& \frac{1}{2} \lim_{T\to \infty}\frac{1}{2\pi i} \int_{\gamma-iT}^{\gamma+iT}\frac{\text{d}\beta}{\beta}e^{\beta \ef} \frac{m}{4\beta\hslash^2} \left( \frac{4}{\omega\beta\hslash}+\omega\beta\hslash \right)\\
    &=&\frac{m}{8\hslash^2}\left( \lim_{T\to \infty}\frac{1}{2\pi i} \int_{\gamma-iT}^{\gamma+iT} \text{d}\beta e^{\beta \ef}  \left( \frac{4}{\omega\beta^3\hslash}+\frac{\omega\hslash}{\beta} \right)\right)\\
    T[\rho]_{SHO}&=& \frac{m\ef^2}{4\omega \hslash^3}+\frac{m\omega}{8\hslash}
\end{eqnarray}
\section{The Kinetic energy contribution from $G^{int}$ \label{gint}in Eq.\eqref{main1D}}
Another essential required for solving the above equation is the multiplication of two Gaussian in our case $G^{SHO}(x,x'';\beta) G^{0}{x'',x';\beta}$
writing the general form
\begin{eqnarray}\label{gaussian multiplication}
    e^{-\alpha_{1}(x-\mu_{1})^2}e^{-\alpha_{2}(x-\mu_{2})^2}
     &=& e^{-(\alpha_1 \mu_{1}^2+\alpha_2 \mu_{2}^2)} e^{-\left(\frac{(\alpha_1 \mu_{1}+\alpha_2 \mu_{2})^2}{\alpha_{1}+\alpha_2}\right)} e^{-(\alpha_{1}+\alpha_2)(x-z)^2}
\end{eqnarray}
 Here $z=\frac{(\alpha_1 \mu_{1}+\alpha_2 \mu_{2})}{\alpha_{1}+\alpha_2}$.
 Similarly, we have to solve another general result for 3rd term
 \begin{eqnarray}\label{general int gaussian}
     \int dx'' (x''-\theta_{1})^2 e^{-\phi(x''-\theta_{2})^2}&=& \int dx'' ((x-\theta_2) + \theta)^2 e^{-\phi(x''-\theta_{2})^2}\\
     &=& \sqrt{\frac{\pi}{\phi}}\left[\frac{1}{2 \phi}+\theta^2 \right]
 \end{eqnarray}
Using eq.\eqref{gaussian multiplication} the multiplication of $G^{SHO}(x,x'';\beta) G^{0}(x'',x':\beta)$ is
\begin{eqnarray*}
    G^{SHO}(x,x'';\beta) G^{0}(x'',x':\beta)&=& e^{-\phi(x''-z)^2} e^{-\alpha(ax^2+ex'^2-bxx')}\\
    \phi&=& (BC+B_{0})\\
    a&=&B^2C^2-B^2+BCB_{0}\\
    e&=& BCB_0 \\
    b&=&2B B_{0}\\
    \alpha &=& 1/\phi\\
    z&=& \frac{Bx+B_{0}x'}{BC+B_{0}}
\end{eqnarray*}
 Solving for the 1st part of $G^{int}$ in Eq.\eqref{main1D}
 \begin{eqnarray*}
     \int dx''  G^{SHO}(x,x'';\beta) G^{0}(x'',x':\beta)&=& A A_{0}\sqrt{\frac{\pi}{BC+B_{0}}} e^{-\alpha(ax^2+ex'^2 -bxx')}\\
     \text{Now  double differentiating $e^{-\alpha(ax^2+ex'^2 -bxx')}$ w.r.t $x$ and $x=x'$}\\
     \left( \alpha^2 (2a-b)^2 x^2-2\alpha a \right)e^{-\alpha.x^2(a+e-b)}\\
       \text{Now  double differentiating $e^{-\alpha(ax^2+ex'^2 -bxx')}$ w.r.t $x'$ and $x=x'$}\\
         \left( \alpha^2 (2e-b)^2 x^2-2\alpha e \right)e^{-\alpha.x^2(a+e-b)}\\
        \text{Now Taking the average}\\
         \frac{1}{2}\left( \alpha^2 (2a-b)^2 x^2-2\alpha a+ \alpha^2 (2e-b)^2 x^2-2\alpha e \right)e^{-\alpha.x^2(a+e-b)}\\
         \text{Integrating \textit{w.r.t} x}\\ 
         \frac{\alpha^2 (2a-b)^2 \pi^{1/2}}{2(\alpha(a+e-b))^{3/2}}- \frac{2 \alpha a \pi^{1/2}}{(a+e-b)^{1/2}}+ \frac{\alpha^2 (2e-b)^2 \pi^{1/2}}{2(\alpha(a+e-b))^{3/2}}-\frac{2 \alpha e \pi^{1/2}}{(a+e-b)^{1/2}}
\end{eqnarray*}
\text{Including the $A A_{0}\sqrt{\frac{\pi}{BC+B_{0}}}$ and using the approximations discussed above which makes $A=A_{0}, B=B_{0}$ we get }\\
\begin{equation}\label{Gint1stafterint}
   \frac{3 (e^{-\beta W_{0}}-1)}{4(C+1)(a+e-b)^{1/2}}\left[ \frac{(2a-b)^2}{2 (a+e-b)}+ \frac{(2e-b)^2}{2(a+e-b) } -2a -2e\right] 
\end{equation}
Using $C=(1+\frac{(\omega \beta \hslash)^2}{2})$
\begin{eqnarray*}
    a+e-b&=& 2 B^2 (\omega \beta \hslash)^2\\
    2a-b&=&3 B^2 (\omega \beta \hslash)^2\\
    2e-b &=&B^2 (\omega \beta \hslash)^2\\
    2e&=&2  B^2 C\\
    2a&=& 2 B^2 \left( 1+ \frac{3(\omega \beta \hslash)^2}{2}\right)
\end{eqnarray*}
By using the above equations, the Eq. \ref{Gint1stafterint} becomes 
\begin{equation}
    -\frac{3(e^{-\beta W_{0}}-1)}{8\sqrt{2}}\frac{m}{\hslash^2}\left( \frac{3\omega \hslash}{4}+ \frac{2}{\omega \hslash \beta^2}\right)
\end{equation}
Hereafter multiplying by the factor of $1/\beta$ before doing the inverse  Laplace  transform
\begin{equation}
    -\frac{3m}{8\sqrt{2} \hslash^2} \left[ \frac{3 \omega \hslash e^{-\beta W_{0}} }{4\beta} + \frac{2  e^{-\beta W_{0}}}{(\omega \hslash \beta^3)} -\frac{3 \omega\hslash}{4 \beta} - \frac{2}{\omega \hslash \beta^3}\right]
\end{equation}
The kinetic energy expression that we got is
\begin{equation}
    T[\rho]_{1}= \frac{3m}{16\sqrt{2}\omega \hslash^3} \left((\ef-W_{0})^2-\ef^2 \right)
\end{equation}
\subsubsection{The kinetic energy contribution from the second part $G^{int}$ of Eq.\eqref{main1D}}
The second part of Eq. \eqref{main1D} is 
\begin{equation}\label{Gint2part}
   = (e^{-\beta W_{0}}-1)\left(B_{0}\int dx''{G}^{SHO}(x, x''; \beta) (x''-x')^2 G^{0}  (x'',x'; \beta) \right) 
\end{equation}
Using the general formula Eq.\eqref{gaussian multiplication} and Eq.\eqref{general int gaussian} we get 

\begin{align*}
    \sqrt{\frac{\pi}{\phi}}\left(\frac{1}{2\phi}+\theta^2 \right)e^-\left({\frac{ax^2+ex'^2-bxx'}{\phi}}\right)\\
    \theta= \frac{B(x-Cx')}{\phi}\\
\text{Simplifying and substituting the value of $\theta$ }\\
    \left( L+M (x-Cx') \right)  e^-\left({\frac{ax^2+ex'^2-bxx'}{\phi}}\right)\\
    \text{here $L$ and $M$ are }\\
    L= \frac{\pi^{1/2}}{2 \phi^{3/2}} ; M= \frac{B^2 \pi^{1/2}}{\phi^{5/2}}  
\end{align*}
Now further we have to double differentiate \textit{w.r.t} $x$ and $x'$
\begin{equation}
    (L+M(x-Cx')^2) e^-\left({\frac{ax^2+ex'^2-bxx'}{\phi}}\right)
\end{equation}
After differentiating \textit{w.r.t} x
\begin{equation}
   [ 2M -4 \alpha M (x-Cx')(2ax-bx')- 2 \alpha a (L+M(x-Cx')^2)+ \alpha^2 (L+M(x-Cx')^2)(2ax-bx')^2] e^{-\alpha( ax^2+ex'^2-bxx')}
\end{equation}
Similarly differentiating \textit{w.r.t} x'
\begin{equation}
    [2MC^2 +4MC\alpha (x-Cx')(2ex'-bx)-2\alpha e (L+M(x-Cx')^2)+\alpha^2 (2ex'-bx)^2(L+M(x-Cx')^2)] e^{-\alpha( ax^2+ex'^2-bxx')}
\end{equation}
After using the approximations $sinh(\omega \hslash \beta)=\omega \hslash \beta $, $cosh(\omega \hslash \beta)= 1+\frac{(\omega \hslash \beta)^2}{2}$ and subsituting the values of $L,M,C, \alpha$ \textit{etc.}. Putting them together in Eq.\eqref{Gint2part} we get
\begin{equation}
    (e^{-\beta W_{0}}-1) \left( \frac{m}{16\sqrt{2} \omega \hslash^3} \left(  \frac{8}{\beta^2}+\frac{43 (\omega \hslash)^2}{8}\right)\right)
\end{equation}
After doing the Laplace transform 
\begin{equation}
    T[\rho]_{2}= \frac{m }{8\sqrt{2} \omega \hslash^3}\left((\ef-W_{0})^2 - \ef^2 \right) 
\end{equation}

\section{Calculation of $\omega_{min}$} 
\label{omwgamincalc}
In order to calculate the $\omega_{min}$ form the kinetic energy  (Eq.\eqref{tot_ke})
\begin{equation}\label{omegaminapp}
    T[\rho] = \frac{m\omega}{8 \hslash}+\frac{m \epsilon_{F}^{2}}{4 \omega \hslash^3}-\frac{5m\epsilon_{F}^{2}}{16\sqrt{2}\omega\hslash^3}+\frac{5 m (\epsilon_{F}-W_{0})^2}{16\sqrt{2} \omega \hslash^3}
\end{equation}
Minimize the $T[\rho]$ \textit{w.r.t} $\omega$
\begin{equation}
    \frac{\partial T[\rho] }{\partial \omega} = \frac{m}{8 \hslash}-\frac{m \epsilon_{F}^{2}}{4 \omega^2 \hslash^3}+\frac{5m\epsilon_{F}^{2}}{16\sqrt{2}\omega^2\hslash^3}-\frac{5 m (\epsilon_{F}-W_{0})^2}{16\sqrt{2} \omega^2 \hslash^3}=0
\end{equation}
The resulting expression for  $\omega_{min}$ is 
\begin{equation}
    \omega_{min} = \frac{\ef}{\hslash}\left( 2- \frac{5W_{0}}{\sqrt{2} \ef}\right)^{1/2}
\end{equation}
\end{appendix}
 \bibliographystyle{apsrev4-1}
 
\bibliography{PRA}
\end{document}


\maketitle

\begin{abstract}
This document provides supplementary information for the main article. It includes additional figures, tables, and detailed descriptions of  methods.
\end{abstract}

\section{Additional Figures and Tables}
\begin{figure}[H]
     \centering
     \begin{subfigure}[b]{0.30\textwidth}
         \centering
         \includegraphics[ angle = 270,width=\textwidth ]{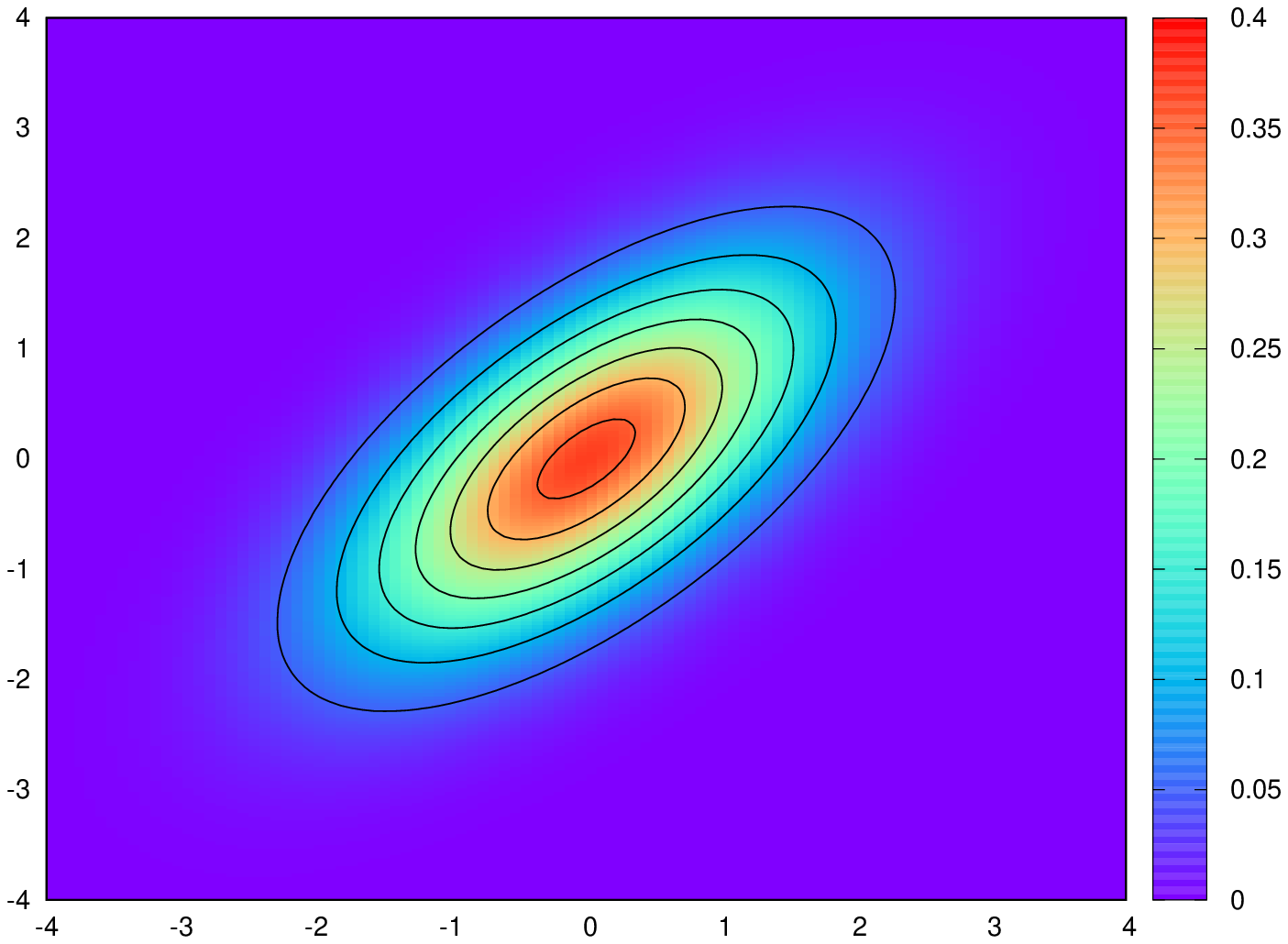}
         \label{fig:y equals x}
     \end{subfigure}
     \hfill
     \begin{subfigure}[b]{0.30\textwidth}
         \centering
         \includegraphics[angle=270, width=\textwidth]{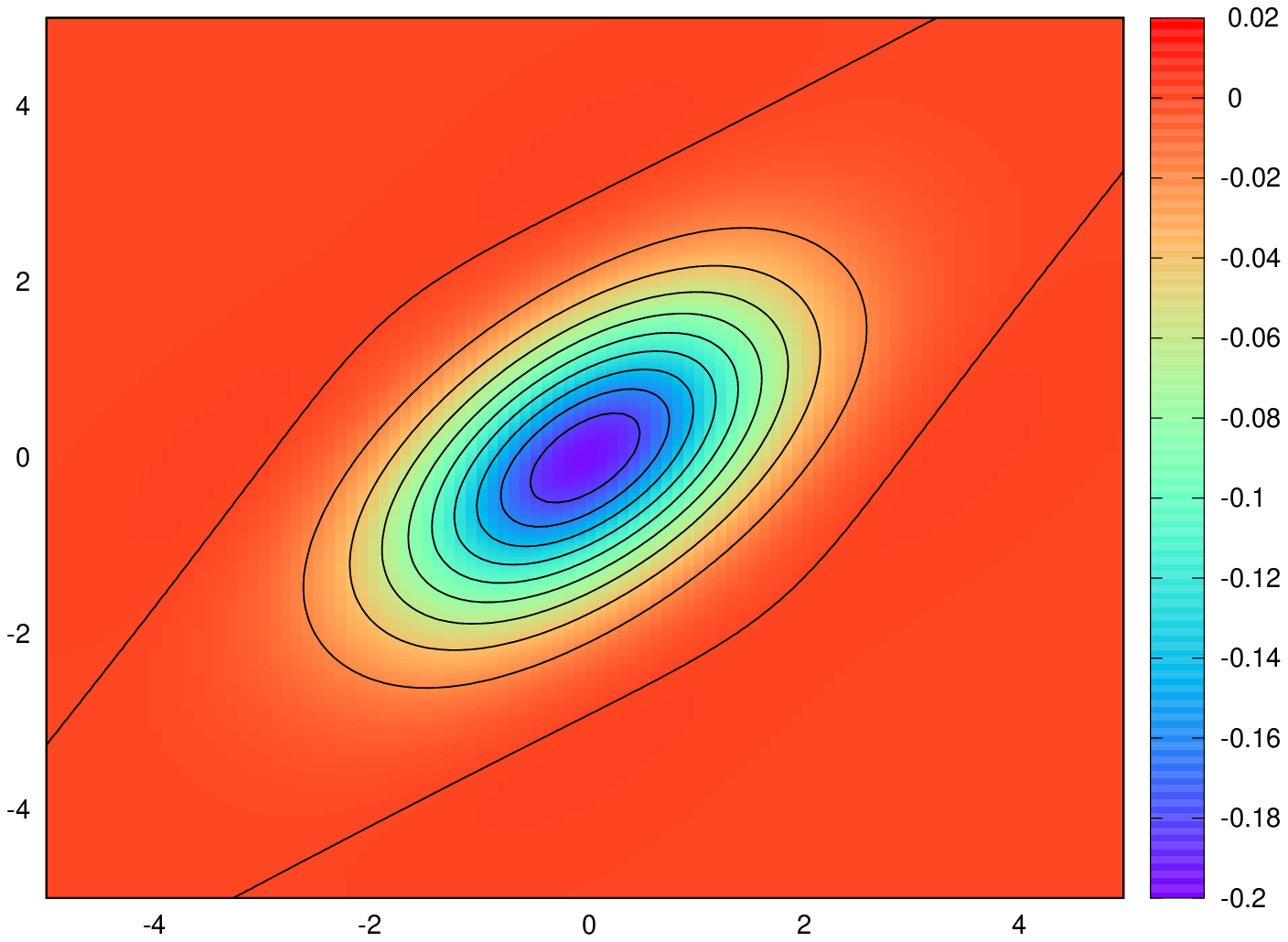}
         \label{fig:three sin x}
     \end{subfigure}
     \hfill
     \begin{subfigure}[b]{0.30\textwidth}
         \centering
         \includegraphics[angle=270, width=\textwidth ]{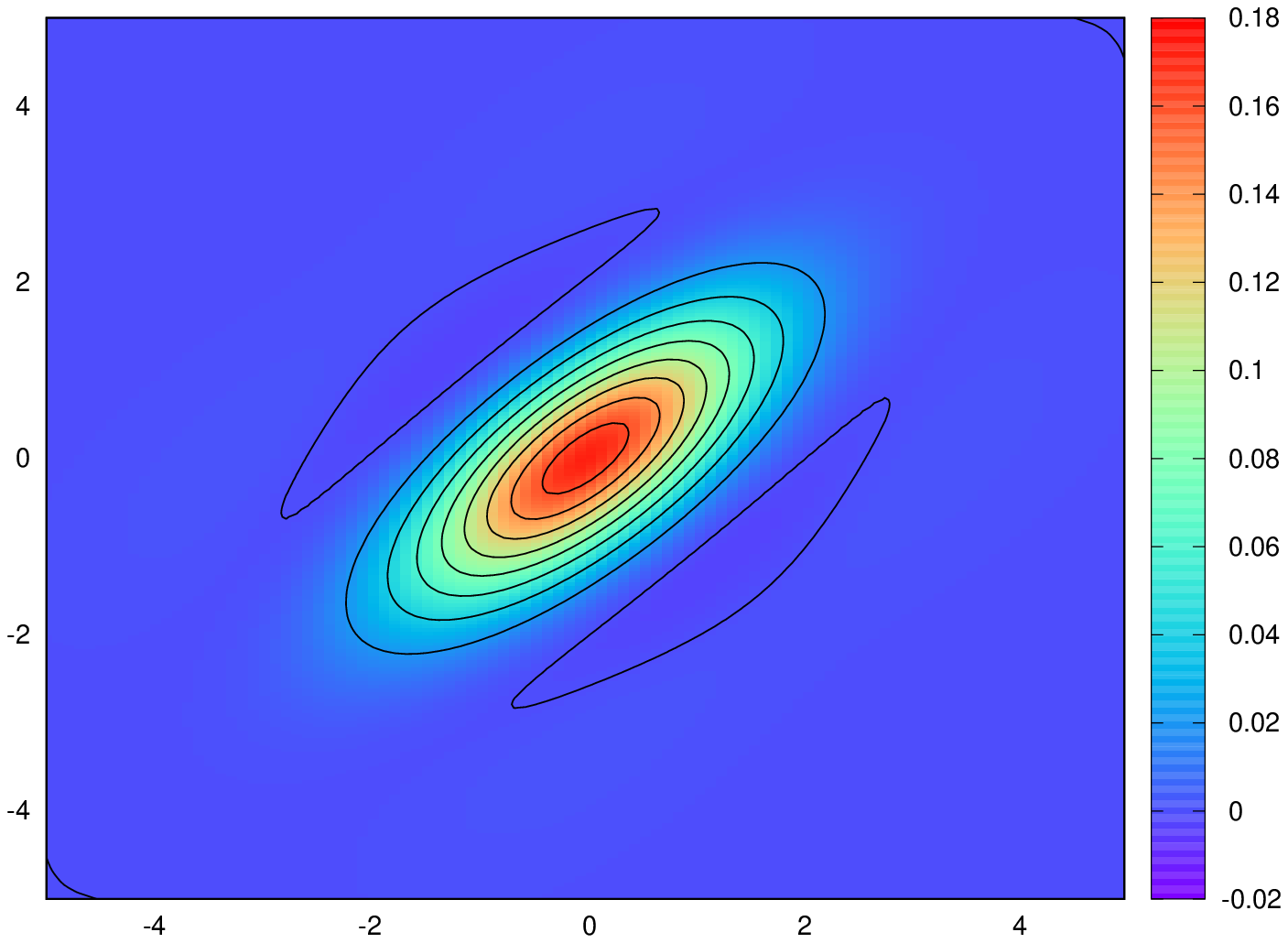}
         \label{fig:five over x}
     \end{subfigure}
        \caption {The contour plot  (a) $G^{SHO}(x,x':\beta)$ is for the Green's function of SHO response with $x$ and $x'$, (b) is the $G^{int}$ \textit{i.e.} interaction term and (c) is the total GF $G(x,x';\beta)$  at $\omega=1$, $\beta=1$}
        \label{fig:three graphs}
\end{figure}

\begin{figure}[H]
     \centering
     \begin{subfigure}[b]{0.30\textwidth}
         \centering
         \includegraphics[ angle = 270,width=\textwidth ]{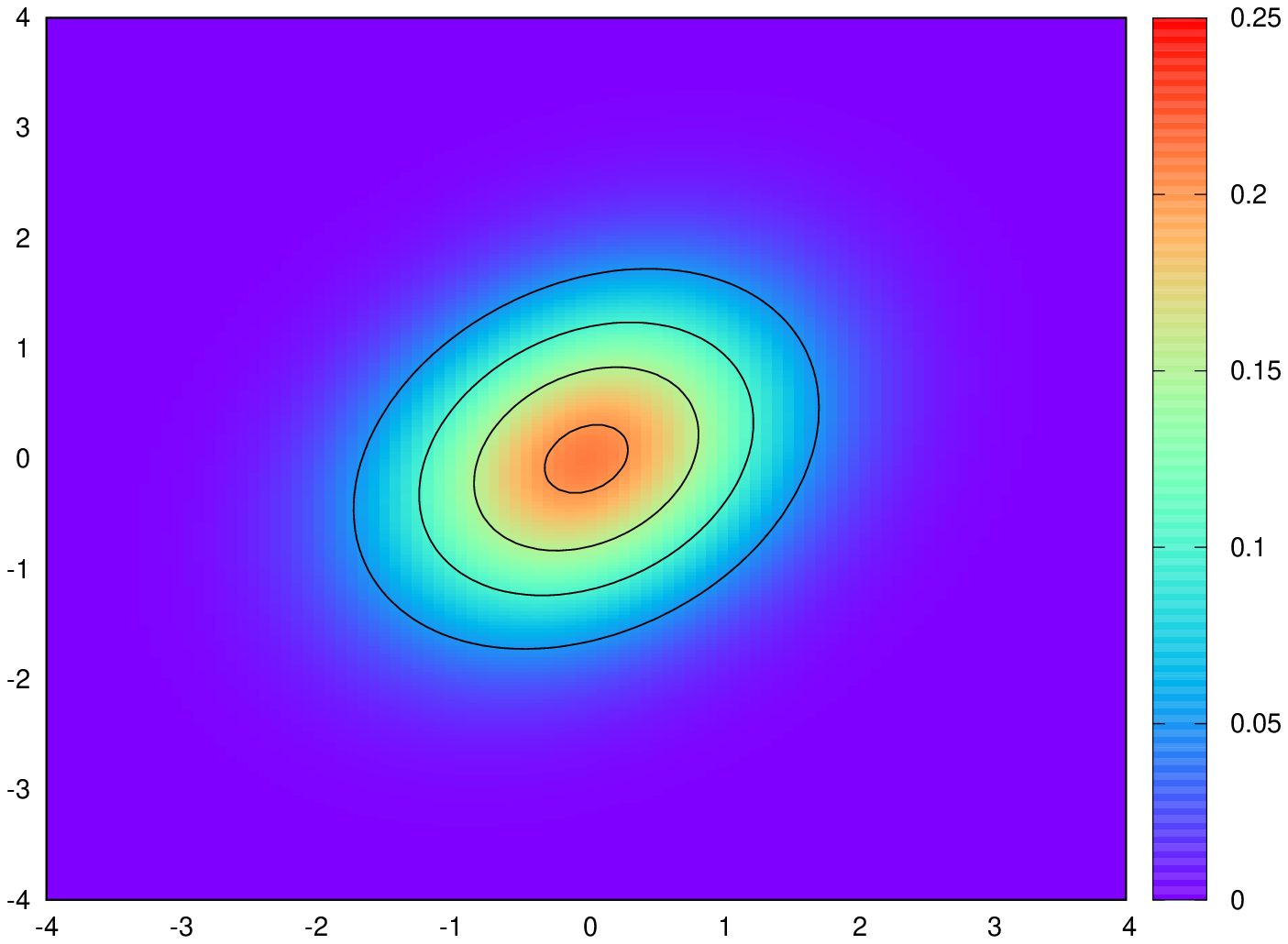}
         \label{fig:y equals x}
     \end{subfigure}
     \hfill
     \begin{subfigure}[b]{0.30\textwidth}
         \centering
         \includegraphics[angle=270, width=\textwidth]{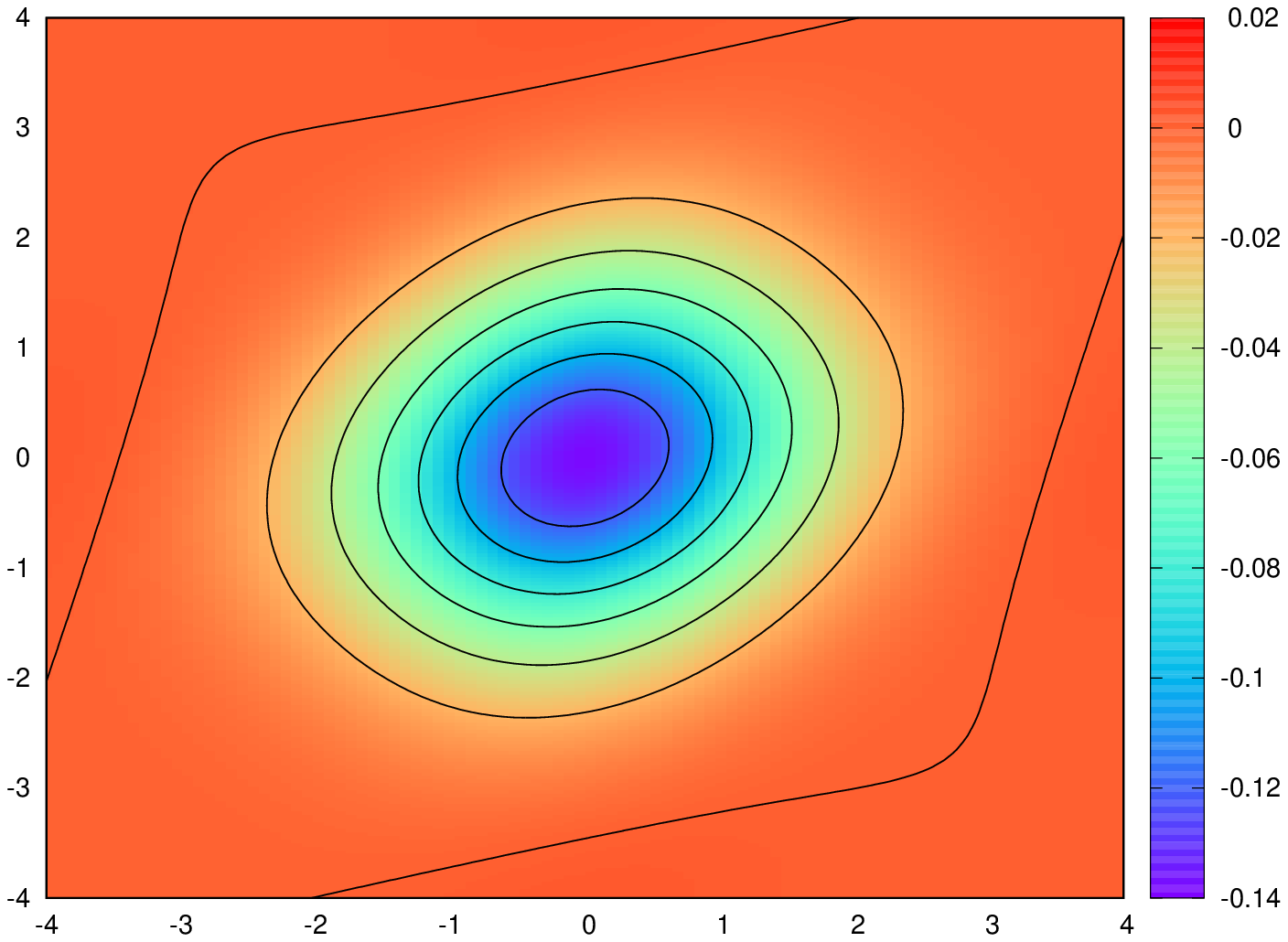}
         \label{fig:three sin x}
     \end{subfigure}
     \hfill
     \begin{subfigure}[b]{0.30\textwidth}
         \centering
         \includegraphics[angle=270, width=\textwidth ]{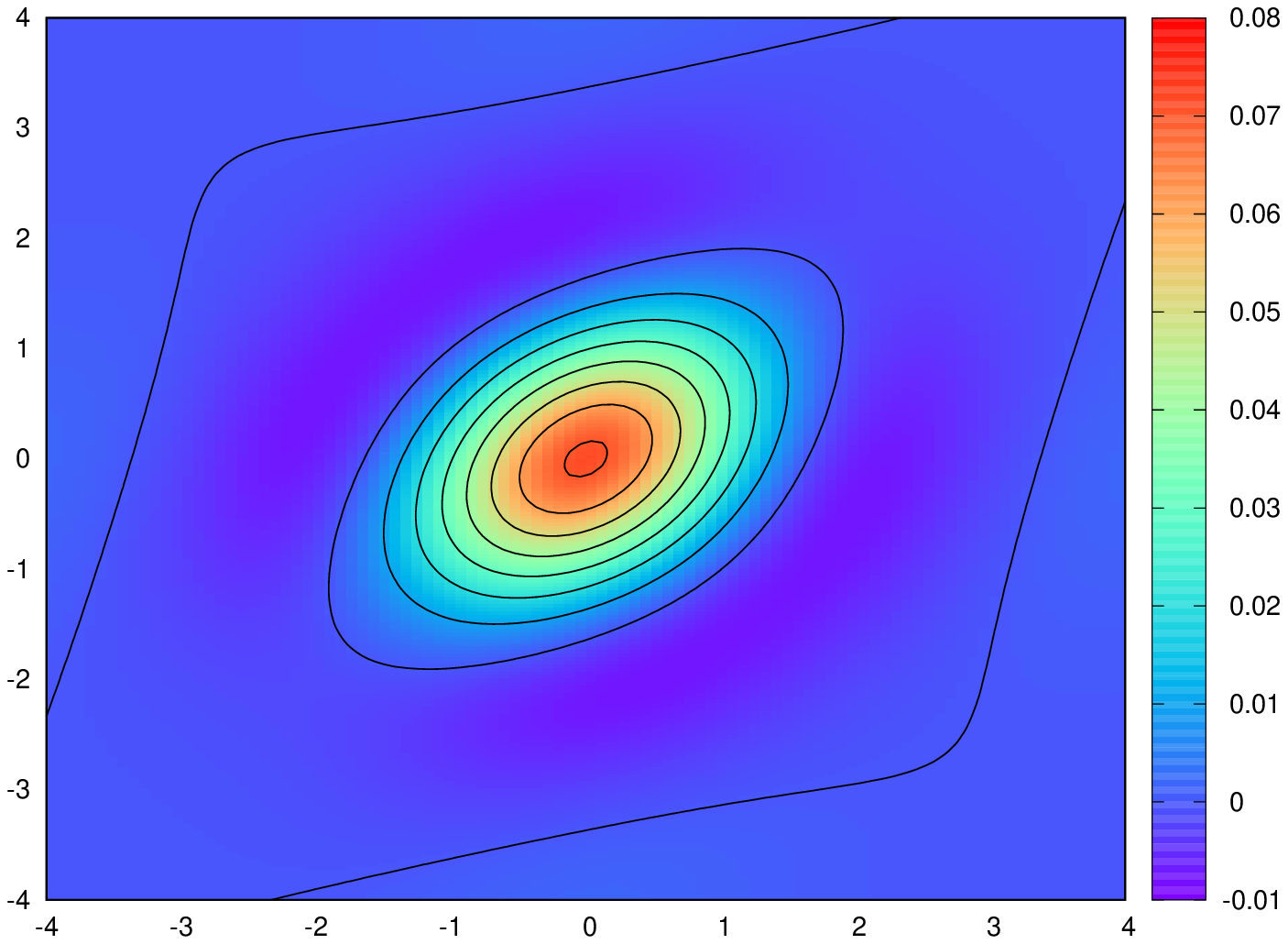}
         \label{fig:five over x}
     \end{subfigure}
        \caption {The contour plot  (a) $G^{SHO}(x,x':\beta)$ is for the Green's function of SHO response with $x$ and $x'$, (b) is the $G^{int}$ \textit{i.e.} interaction term and (c) is the total GF $G(x,x';\beta)$  at $\omega=1$, $\beta=2$}
        \label{fig:three graphs}
\end{figure}
\begin{figure}[H]
     \centering
     \begin{subfigure}[b]{0.30\textwidth}
         \centering
         \includegraphics[ angle = 270,width=\textwidth ]{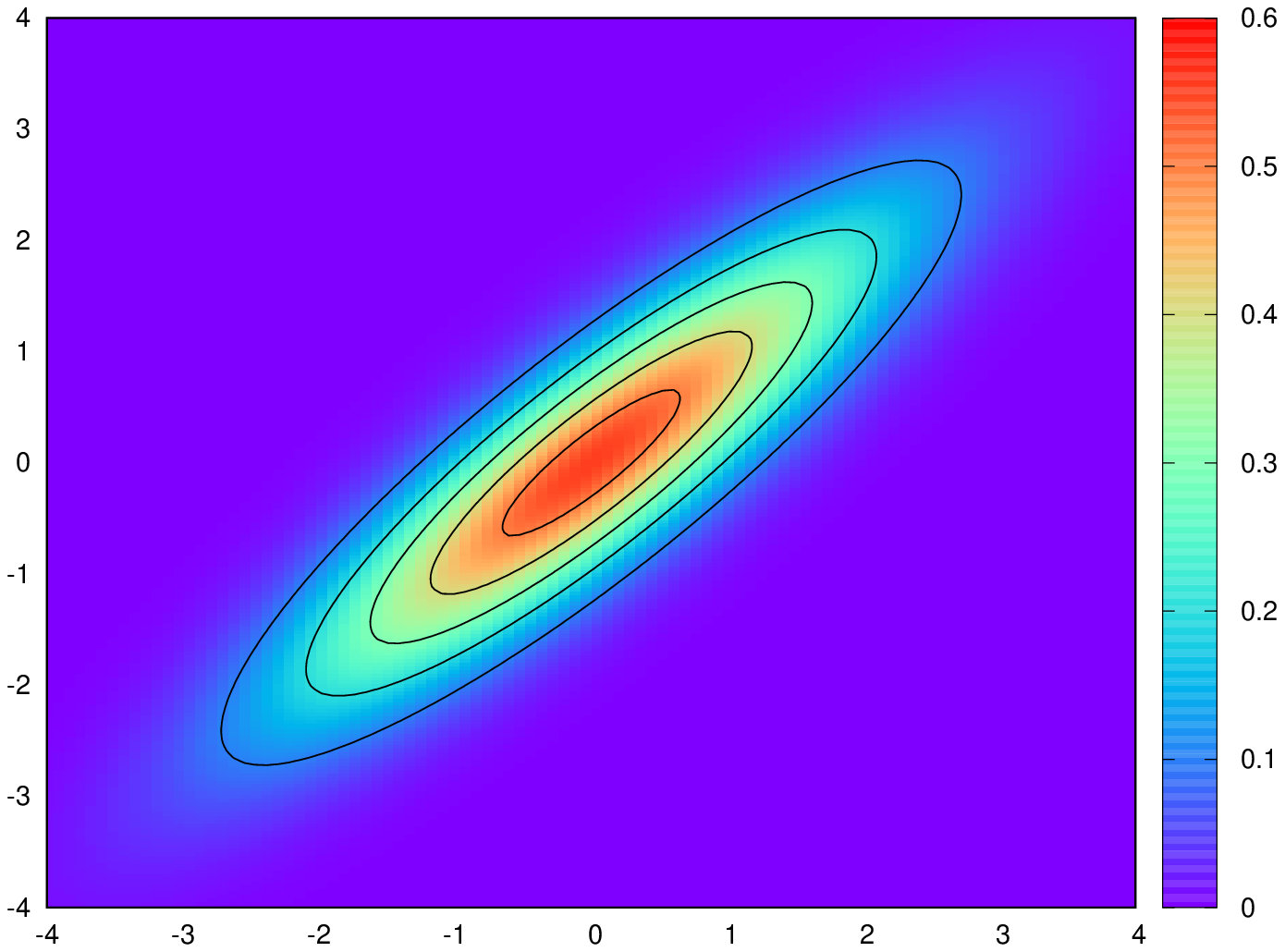}
         \label{fig:y equals x}
     \end{subfigure}
     \hfill
     \begin{subfigure}[b]{0.30\textwidth}
         \centering
         \includegraphics[angle=270, width=\textwidth]{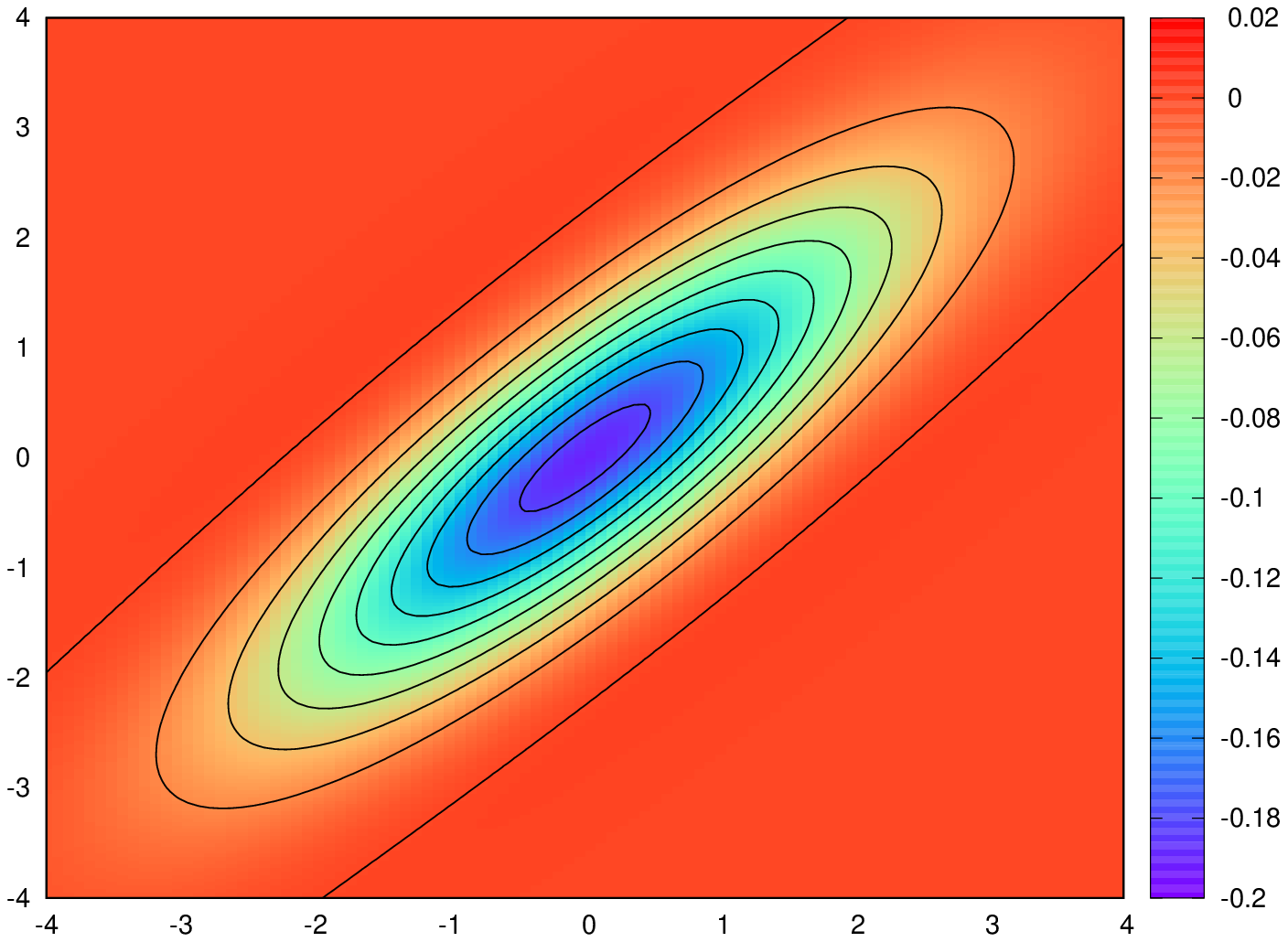}
         \label{fig:three sin x}
     \end{subfigure}
     \hfill
     \begin{subfigure}[b]{0.30\textwidth}
         \centering
         \includegraphics[angle=270, width=\textwidth ]{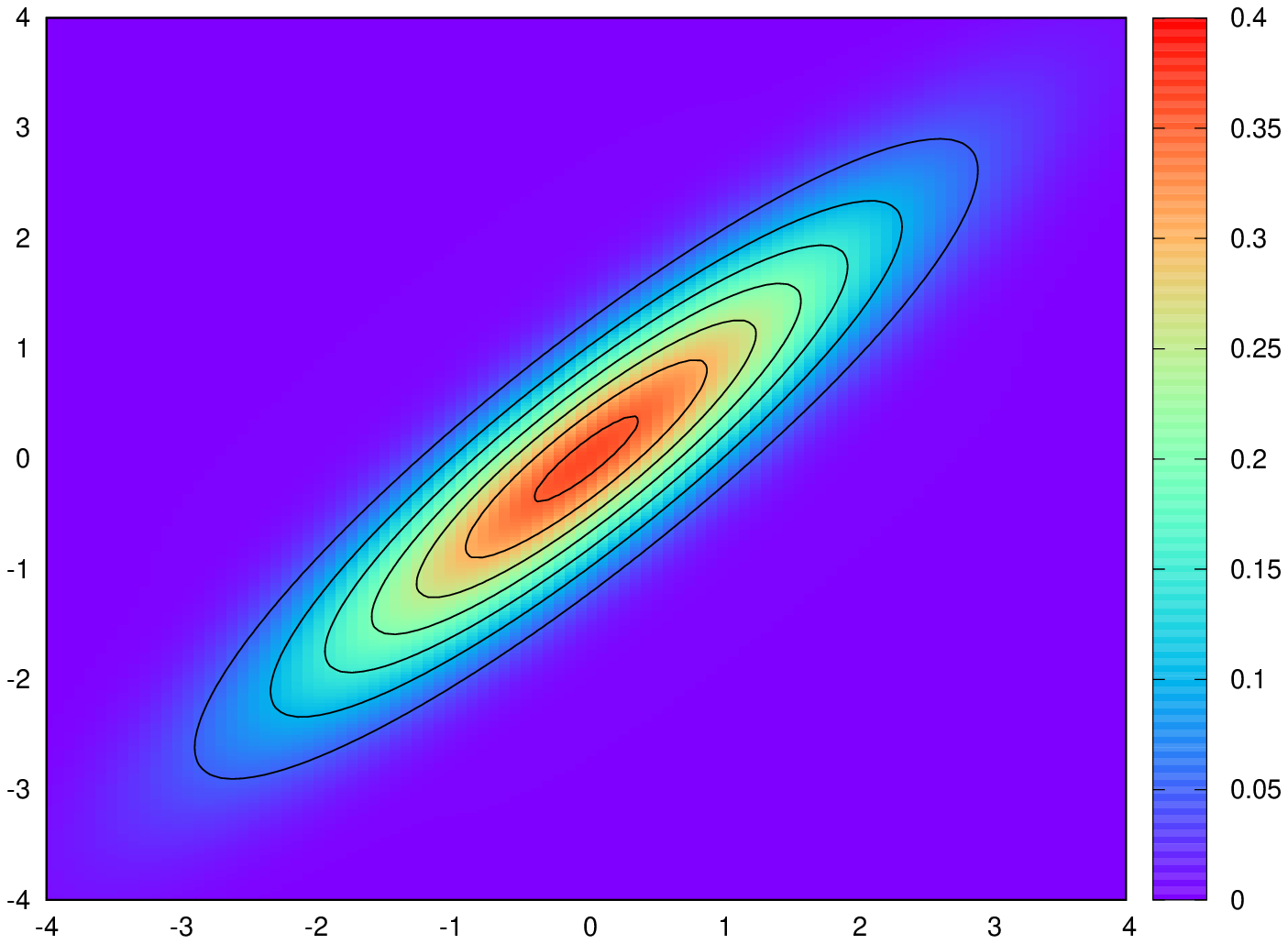}
         \label{fig:five over x}
     \end{subfigure}
        \caption {The contour plot  (a) $G^{SHO}(x,x':\beta)$ is for the Green's function of SHO response with $x$ and $x'$, (b) is the $G^{int}$ \textit{i.e.} interaction term and (c) is the total GF $G(x,x';\beta)$  at $\omega=1$, $\beta=0.5$}
        \label{fig:three graphs}
\end{figure}

\subsection{contourlines information}

\begin{table}[h!]
    \centering
    \begin{tabular}{|c|c|c|c|}
        \hline
        Contour line & $G^{SHO}$ & $G^{int}$ & $G$ \\ \hline
        $\omega=0.5$ & 0.005 & 0.002 & 0.002 \\ \hline
        $\omega=1$ & 0.0005 & 0.0002 & 0.0005 \\ \hline
        $\omega=2$ & 0.000005 & 0.000002 & 0.000005 \\ \hline
    \end{tabular}
    \caption { The step size for contour lines for $G$, $G^{int}$ and $G^{SHO}$ for different $\omega $ values at $\beta=10$}
    \label{tab:simple_table}
\end{table}